\documentclass[twocolumn]{aastex631}
\usepackage{amsmath}
\usepackage{natbib}

\shorttitle{Dark snails}
\shortauthors{Gilman et al.}

\begin{document}
	
	\newcommand{\vdag}{(v)^\dagger}
	\newcommand\aastex{AAS\TeX}
	\newcommand\latex{La\TeX}
	
	\title[dark subhalos and the snail]{Dark Galactic subhalos and the Gaia snail}
	
	\author[0000-0002-5116-7287]{Daniel Gilman}
	\affiliation{Department of Astronomy and Astrophysics, University of Chicago, Chicago, IL 60637, USA}
	\affiliation{Brinson Prize Fellow}
	\affiliation{Department of Astronomy and Astrophysics, University of Toronto, 50 St. George Street, Toronto, ON, M5S 3H4, Canada}
	
	\author[0000-0001-6855-442X]{Jo Bovy}
	\affiliation{Department of Astronomy and Astrophysics, University of Toronto, 50 St. George Street, Toronto, ON, M5S 3H4, Canada}
	
	\author[0000-0002-6411-8695]{Neige Frankel}
	\affiliation{Department of Astronomy and Astrophysics, University of Toronto, 50 St. George Street, Toronto, ON, M5S 3H4, Canada}
	\affiliation{Canadian Institute for Theoretical Astrophysics, University of Toronto, 60 St. George Street, Toronto, ON M5S 3H8, Canada}
	
	\author[0000-0001-5501-6008]{Andrew Benson}
	\affiliation{Carnegie Institution for Science, 813 Santa Barbara Street, Pasadena, CA 91101, USA}
	\correspondingauthor{Daniel Gilman}
	\email{gilmanda@uchicago.edu}
	\begin{abstract}
		Gaia has revealed a clear signal of disequilibrium in the solar neighborhood in the form of a spiral (or snail) feature in the vertical phase-space distribution. We investigate the possibility that this structure emerges from ongoing perturbations by dark $\left(10^{6} M_{\odot} - 10^8 M_{\odot}\right)$ Galactic subhalos. We develop a probabilistic model for generating subhalo orbits based on a semianalytic model of structure formation, and combine this framework with an approximate prescription for calculating the response of the disk to external perturbations. We also develop a phenomenological treatment for the diffusion of phase-space spirals caused by gravitational scattering between stars and giant molecular clouds, a process that erases the kinematic signatures of old ($t \gtrsim 0.6$ Gyr) events. Perturbations caused by dark subhalos are, on average, orders of magnitude weaker than those caused by luminous satellite galaxies, but the ubiquity of dark halos predicted by cold dark matter makes them a more probable source of strong perturbation to the dynamics of the solar neighborhood. Dark subhalos alone do not cause enough disturbance to explain the Gaia snail, but they excite fluctuations of $\sim 0.1-0.5 \ \rm{km} \ \rm{s^{-1}}$ in the mean vertical velocity of stars near the Galactic midplane that should persist to the present day. Subhalos also produce correlations between vertical frequency and orbital angle that could be mistaken as originating from a single past disturbance. Our results motivate investigation of the Milky Way's dark satellites by characterizing their kinematic signatures in phase-space spirals across the Galaxy. 
	\end{abstract}
	\keywords{
		\href{http://astrothesaurus.org/uat/353}{Dark matter (353)}; \href{http://astrothesaurus.org/uat/1596}{Stellar dynamics (1596)}; 
		\href{http://astrothesaurus.org/uat/1594}{Galaxy stellar disks (1594)};
		\href{http://astrothesaurus.org/uat/2360}{Gaia (2360)}
	}
	
	\section{Introduction}
	
	The Gaia mission has revolutionized research in Galactic dynamics through precise kinematic measurements of a billion stars \citep{Gaia++16}. The exquisite data obtained through this survey \citep[e.g.][]{GaiaDR2,GaiaDR3} reveals in unprecedented detail an unexpected phenomenon discovered by \citet{Widrow++12}: the distribution function of stars in the solar neighborhood, the portion of the Milky Way $\sim 8 \ \rm{kpc}$ from the Galactic center and confined to $|z| \lesssim 2 \ \rm{kpc}$ away from the Galactic midplane, exhibits clear deviations from dynamic equilibrium. Correlations between the vertical velocity and height above the Galactic midplane manifest as a prominent spiral in the vertical ($z$-direction) phase-space distribution \citep{Antoja++18,BennettBovy19}. 
	
	The phase-space spiral detected by Gaia, sometimes referred to as the Gaia snail, reflects the response of the Galactic disk to one or several perturbations to the dynamics of the solar neighborhood. The evolution of a perturbation into a spiral pattern in the distribution function occurs as a result of phase mixing \citep[e.g.][]{Tremaine99}: because the frequency of vertical oscillations near the Galactic midplane changes with height, over time an initial displacement of test particles (stars) from their equilibrium trajectories winds into a spiral. The properties of emergent spirals depend on the gravitational potential of the Galactic disk, which determines how quickly an initial perturbation becomes sheared and distorted \citep[e.g.][]{Widrow++12,Widmark++21,Banik++22,Banik++23}. 
	
	The properties of phase spirals also depend on the source of the perturbation. Due to its close passage to the solar neighborhood in the past gigayear, the Sagittarius dwarf galaxy, or a massive satellite with similar properties, was regarded as the most probable culprit for the dynamic perturbation that spawned the Gaia snail \citep{Gomez++13,Laporte++18,Laporte++19,BlandHawthorn++19,BennettBovy21,BennettBovy22}. However, further investigation revealed problems with this hypothesis. Although Sagittarius appears to cause a detectable perturbation, models that include only Sagittarius struggle to explain the amplitude of the perturbation measured by Gaia, even when assuming it has retained most of its mass since accretion onto the Milky Way. 
	
	Assuming a single event caused the perturbation, it is possible to unwind the snail to determine when the perturbation occurred \citep[e.g.,][]{Binney++18,BlandHawthorn++19,Widmark++22,Gandhi++22,DarraghFord++23}. However, the Gaia snail defies association with a single dynamic perturbation time, and instead suggests multiple events occurred over the past $\sim 0.5 \ \rm{Gyr}$ \citep{Frankel++23,Antoja++23}. The snail exhibits various other complex features, such as correlations between perturbation strength, timing, and angular momentum, suggesting that an interplay between numerous factors contribute to the observed phenomenon \citep{Binney++18,LiWidrow21,Hunt++22,Gandhi++22,Frankel++23,LiWidrow23,Grand++23,Hunt++24,Frankel++24}. We note, however, that \cite{DarraghFord++23} show that a single perturbation can produce a complex signal when accounting for the 3D configuration of the system. The transition from a single-arm to doubled-armed spiral seen in the Gaia data is similarly difficult to explain if one assumes a single event caused the snail \citep{Hunt++22}. Various other sources of perturbation, such as the Galactic bar or the Galaxy's spiral arms \citep{Monari++15,Hunt++18,Khoperskov++19}, could contribute to the out-of-equilibrium properties of the local phase space, although, considered individually, they struggle to explain all aspects of the data. 
	
	As shown by \citet{Tremaine++23}, an ensemble of small perturbations, rather than one large perturbation, can also give rise to coherent phase spirals. Cold dark matter (CDM) provides a source of these small perturbations in the form of dark-matter subhalos. According to CDM, most galaxies reside inside of halos with masses at infall $\gtrsim 10^{8} M_{\odot}$. Below $10^{8} M_{\odot}$, the fraction of halos that retain enough stars to form a detectable galaxy plummets toward zero  \citep[e.g.][]{Nadler++20}. The abundance of halos scales approximately as the inverse of the halo mass. Since at least one halo, the host of the Sagittarius dwarf galaxy, has come close enough to the solar neighborhood to affect the vertical dynamics, many more lower-mass halos could have passed in even closer proximity. Therefore, the question is not whether these more ubiquitous low-mass halos can cause perturbations to the Gaia snail, but how this signal manifests, and whether it is detectable. 
	
	The idea that the Milky Way satellite population could affect the properties of galactic disks has existed since long before Gaia launched into space \citep[e.g.][]{Quinn++93}. However, most analyses consider their effects on larger scales than those relevant for understanding the detailed properties of the Gaia snail, and only consider the impact of the most-massive perturbers, objects which should host detectable galaxies. For example, \citet{Kazantzidis++08} performed N-body simulations of $\sim 10^{10} M_{\odot}$ objects interacting with a stellar disk, and identified several morphological features across the disk that form as a result of these interactions. \citet{MoetazedianJust16} showed that the vertical heating across the Milky Way's disk is dominated by the most-massive subhalos, a result confirmed by \citet{Grand++16}. \citet{GarciaConde++24} performed high-resolution simulations of the Galactic disk evolving in the presence of several sources of perturbation, including subhalos, and identified correlations between breathing and bending modes in the disk and the passage of suhbalos. Their results were consistent with the analysis presented by \citet{Chequers++18}, who found that substructure can excite existing bending modes in galactic disks. Both of these analyses considered relatively massive $10^8 M_{\odot} - 10^9 M_{\odot}$ halos, comparable to the present-day mass of Sagittarius \citep{VasilievBelokurov20}. Earlier, \citet{Feldmann++15} and \citet{Buschmann++18} simulated the effects of halos with mass $\sim 10^{9} M_{\odot}$. They predicted perturbations to the velocities of stars of order $1 \ \rm{km} \ \rm{s^{-1}}$, and estimated the effects of halos with masses as low as $10^7 M_{\odot}$ could be detectable with Gaia. 
	
	The high resolution in space, velocity, and particle mass needed to resolve the detailed properties of the Gaia snail has limited further investigation of how the Galactic disk responds to disturbances by low-mass satellites. Numerical simulations with billions of particles \citep[e.g.][]{Hunt++21,BennettBovy22} can achieve the precision required to study the properties of transient density waves through the galaxy. However, the computational cost of running these simulations limits their utility for calculating the statistical properties of weak perturbations from low-mass satellites, which involves considering many possible realizations of perturber orbits, and different models for the gravitational potential of the galaxy. 
	
	In this work, we directly address the question of whether ongoing perturbations by low-mass halos ($m < 10^{8} M_{\odot}$) can reproduce features of the Gaia snail. In particular, the multiple perturbations by dark halos could plausibly explain why the snail appears more consistent with numerous disturbances, rather than originating from a single event. This work builds on analyses by \citet{Feldmann++15} and \citet{Buschmann++18}, who considered perturbations by a single passing subhalo and argued that the imprints of a population of these objects could impart observable imprints on the kinematics of stars. In our treatment of this problem, we also implement a phenomenological model of diffusion based on \citet{Tremaine++23}, a process that occurs as a consequence of gravitational scattering between stars and giant molecular clouds. To calibrate a model for dark-matter substructure we rely on a semianalytic model of structure formation that accurately predicts the evolution of subhalo populations in Milky Way-like galaxies orders of magnitude faster than N-body simulations. Using an approximate one-dimensional model for the vertical dynamics of stars near the Galactic disk, we study the response of stellar orbits to gravitational encounters with dark subhalos at arbitrarily high spatial and velocity resolution in the phase-space distribution. In particular, we will examine, across multiple realizations of possible subhalo orbits, how the phase spirals appear in frequency-angle coordinates, how the strength of perturbations depend on the abundance, tidal evolution, and mass definition of subhalos, and how various summary statistics, such as the vertical number count asymmetry and mean vertical velocity, respond to the bombardment of the Galactic disk by these objects. 
	
	This paper is organized as follows: In Section \ref{sec:forwardmodeling} we describe our model for the calculation of the perturbed phase-space distribution, and our prescription for generating realistic substructure populations. In Section \ref{sec:signatures} we examine the statistical properties of subhalo encounters with the solar neighborhood, and illustrate several examples of how these structures manifest in the vertical phase-space distribution. We summarize our main results and provide concluding remarks in Section \ref{sec:conclusions}. 
	
	We have released the code used for this analysis in an open-source python package, {\tt{darkspirals}}\footnote{\url{https://github.com/dangilman/darkspirals}}. Several example notebooks illustrate the core functionality of the package, and effort has been made to provide clear documentation. Orbit integration is performed with {\tt{galpy}}\footnote{\url{https://github.com/jobovy/galpy}} \citep{Bovy15}. \\
	
	\section{Forward modeling the formation and diffusion of phase-space spirals}
	\label{sec:forwardmodeling}
	In this section, we start in Section~\ref{ssec:modelspirals} by describing how we calculate the observed $z - v_z$ phase-space distribution subject to ongoing perturbations from external sources following the methods presented by \citet{BennettBovy21} (see also \citealt{Kalnajs73}). In Appendix \ref{app:A}, we examine the accuracy of the framework discussed in this section for predicting the properties of phase-space spirals subject to multiple perturbations. In Section \ref{ssec:modeldiffusion}, we describe a phenomenological model for diffusion, which erases signatures of old perturbations. Section \ref{ssec:satellites} discusses how we model the population of dark and luminous satellites, including our prescription for generating subhalo orbits and masses. 
	
	\subsection{A forward model for nonequilibrium dynamics}
	\label{ssec:modelspirals}
	\begin{figure}
		\centering
		\includegraphics[trim=0cm 0cm 0cm
		0.cm,width=0.45\textwidth]{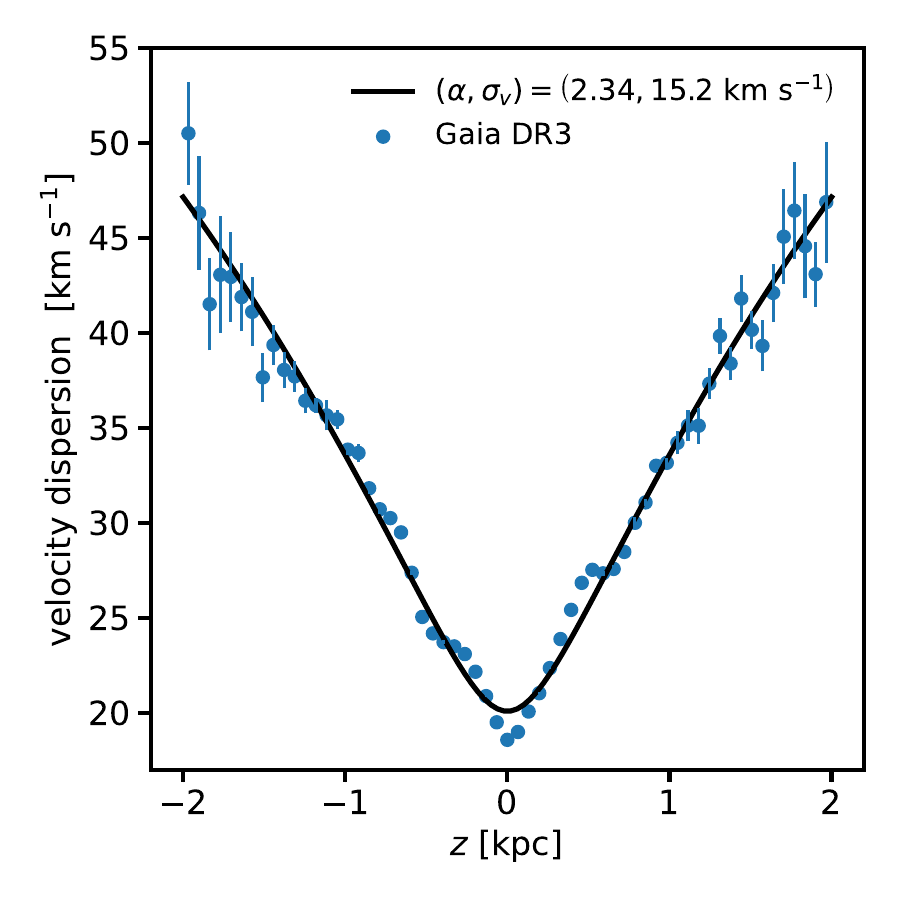}
		\caption{\label{fig:vdisfit} The velocity dispersion predicted by the distribution function given in Equation \ref{eqn:liwidrowdf} (black curves) fit to the velocity dispersion of stars in the solar neighborhood measured by Gaia (blue points). We have adjusted the original formula presented by \citet{LiWidrow21} to use the action and vertical frequency, $J_z \nu$, in place of the orbital energy $E_z$, and fit the parameters $\alpha$ and $\sigma_v$ to the data.}
	\end{figure}
	We consider a scenario in which particles orbit in a one-dimensional gravitational potential, an approximation of the gravitational field in the immediate vicinity ($|z| \sim \mathcal{O}\left(1 \right) \ \rm{kpc}$) of the Galactic midplane. In dynamic equilibrium, the stellar distribution function, $f_{\rm{eq}}\left(J\right)$, and the Hamiltonian, $H_{\rm{eq}}\left(J\right)$, are only functions of the vertical action $J \equiv J_z$. Because perturbations are small, we work in the action-angle coordinates $(J,\theta) \equiv (J_{\rm eq},\theta_{\rm{eq}})$ computed in the equilibrium potential. The perturbed Hamiltonian is then 
	\begin{equation}
		H\left(J, \theta\right) = H_{\rm{eq}}\left(J\right) + \Psi\left(J, \theta\right),
	\end{equation}
	where $\Psi\left(J,\theta\right)$ represents the potential from an external perturbation. The first of Hamilton's equations gives 
	\begin{eqnarray}
		\label{eqn:ham1}
		\frac{d \theta}{d t}  &=& \frac{\partial}{\partial J}\left(H_{\rm{eq}} + \Psi \right) = \Omega_{\rm{eq}}\left(J\right) + \frac{\partial \Psi}{\partial J}\,.
	\end{eqnarray} 
	The second of Hamilton's equations gives
	\begin{eqnarray}
		\label{eqn:ham2}
		\frac{d J}{d t}  &=& -\frac{\partial}{\partial \theta}\left(H_{\rm{eq}} + \Psi \right)  \\
		\nonumber &=& -\frac{\partial \Psi}{\partial \theta} \\
		\nonumber &=& -\frac{\partial t}{\partial \theta} \cdot \frac{\partial z}{\partial t} \cdot \frac{\partial \Psi\left(J,  \theta\left(t\right), t\right)}{\partial z}\\
		\nonumber &=& \frac{v_z}{\left(\partial t/\partial \theta\right)^{-1}} F_{\rm{z}}\left(J, \theta\left(t\right), t\right)
	\end{eqnarray}
	where in the third line we have applied the chain rule, and in the last we have replaced the derivatives with the vertical velocity $v_z$ and an external force $F_z$.
	
	For what follows, we assume stars orbit in an equilibrium gravitational potential, $\Phi_{\rm{eq}}$, which we take to be the {\tt{MWPotential2014}} potential in {\tt{galpy}} \citep{Bovy15}. For small perturbations $|\partial \Psi / \partial z| << |\partial \Phi_{\rm{eq}}/\partial z|$, we can approximate the change in the vertical action by integrating Equation \eqref{eqn:ham2} backwards along the trajectory of an unperturbed orbit
	\begin{equation}
		\label{eqn:dji}
		\Delta J = \int_{0}^{-T} \frac{v_z\left(t^{\prime}\right)}{\Omega} F_{\rm{z}}\left(J, \theta\left(t^{\prime}\right), t^{\prime}\right) \mathrm{d}t^{\prime},
	\end{equation}
	where we have written $\left(\partial t/\partial \theta\right)^{-1} = \Omega = \Omega_{\rm{eq}}\left(J\right)$ for the unperturbed orbit\footnote{$|\partial \Psi / \partial J| << |H_{\rm{eq}} / \partial J|$ for small perturbations so $\Omega \approx \Omega_{\rm{eq}}\left(J\right)$.}. We integrate until $T = 1.2 \ \rm{Gyr}$, encompassing the range of times during which a perturbation caused by a satellite will persist until the present day, subject to diffusion through gravitational encounters with giant molecular clouds (see Section \ref{ssec:modeldiffusion}). 
	
	In this work, we consider a scenario in which the vertical phase-space distribution is subjected to many small external forces $F_{z,i}$ which, from Equation \eqref{eqn:dji}, cause a change in the action $\Delta J_{i}$. The new vertical action is $J = J_{\rm{eq}} + \sum \Delta J_{i}$, where $J_{\rm{eq}}$ is the action of the unperturbed orbit. To calculate the resulting phase-space distribution, we use the rational linear distribution function introduced by \citet{LiWidrow21}, but written in terms of the vertical action using the substitution $E_z = J\,\nu$, where $\nu =\partial^2 \Phi_{\rm{eq}}\left(R,z\right) / \partial^2 z \big \vert_{\left(R,z\right)=\left(R_0,0\right)}$ is the vertical frequency \citep{Binney++11}:
	\begin{equation}
		\label{eqn:liwidrowdf}
		f\left(J\right) = n_0 \left(1 + \frac{J \nu}{\alpha \sigma_z^2}\right)^{-\alpha}.
	\end{equation}
	This distribution function corresponds to a superposition of isothermal distribution functions with a continuous change of velocity dispersion with distance from the Galactic midplane. Here, $\sigma_z$ represents a characteristic velocity scale that approaches the velocity dispersion of an isothermal distribution function as $\alpha \rightarrow \infty$. When modeling the local vertical phase-space distribution, we evaluate the vertical frequency at $R_0 = 8.178 \ \rm{kpc}$, the distance between the Sun and the Galactic center \citep{GravityCollab}. As in \citet{BennettBovy21}, we evaluate the equilibrium distribution function in Equation \ref{eqn:liwidrowdf} at the perturbed coordinates $J = J_{\rm{eq}} + \sum_{i} \Delta J_i$. Throughout this paper, we will use notation $f\left(z,v_z\right)$ in place of $f\left(J\left(z,v_z\right)\right)$. 
	\begin{figure*}
		\centering
		\includegraphics[trim=0.5cm 0.25cm 0.5cm
		0.5cm,width=0.48\textwidth]{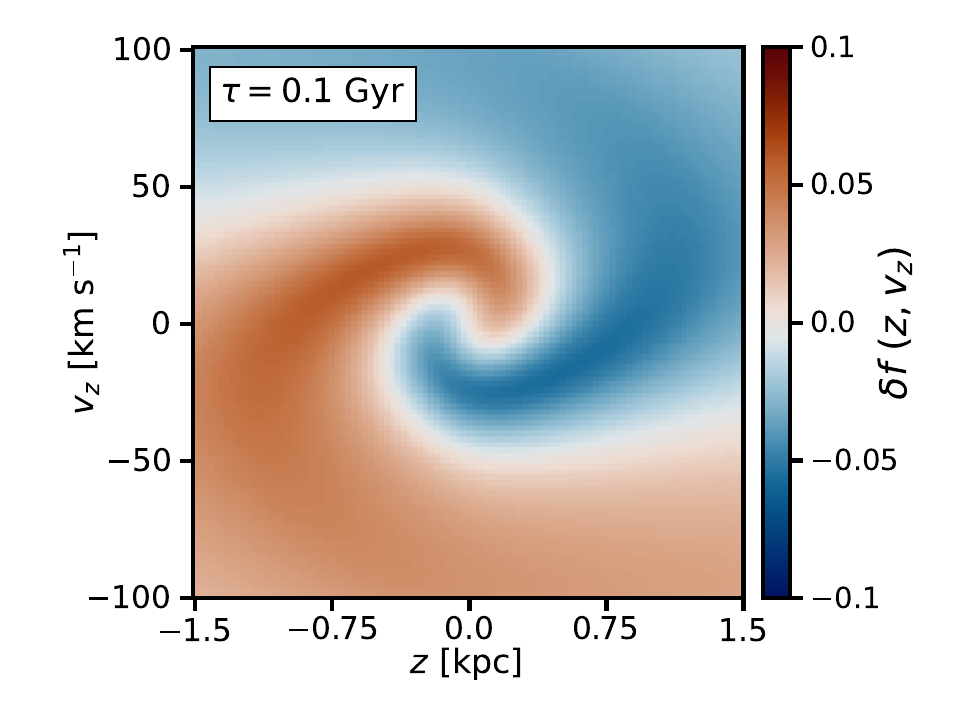}
		\includegraphics[trim=0.5cm 0.5cm 0.5cm
		0.5cm,width=0.48\textwidth]{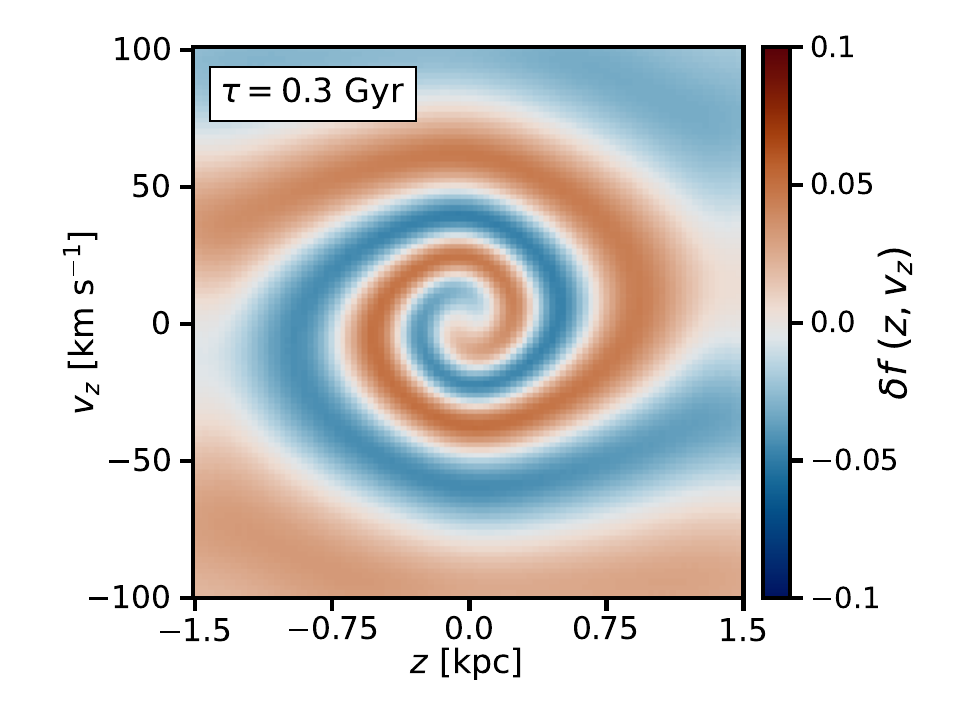}
		\includegraphics[trim=0.5cm 0.25cm 0.5cm
		0.5cm,width=0.48\textwidth]{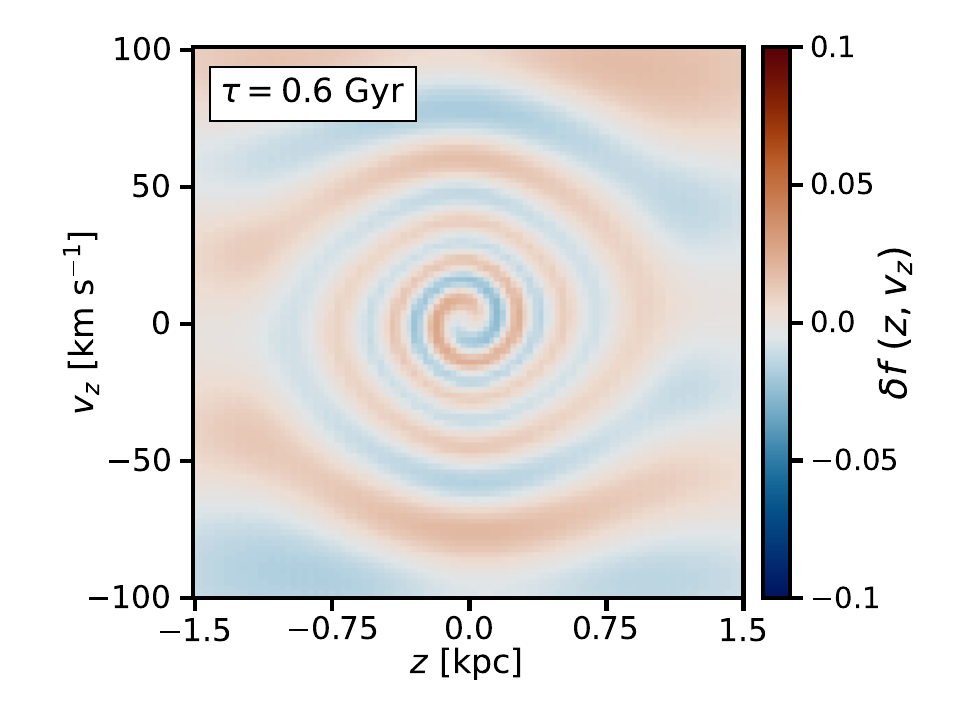}
		\includegraphics[trim=0.5cm 0.5cm 0.5cm
		0.5cm,width=0.48\textwidth]{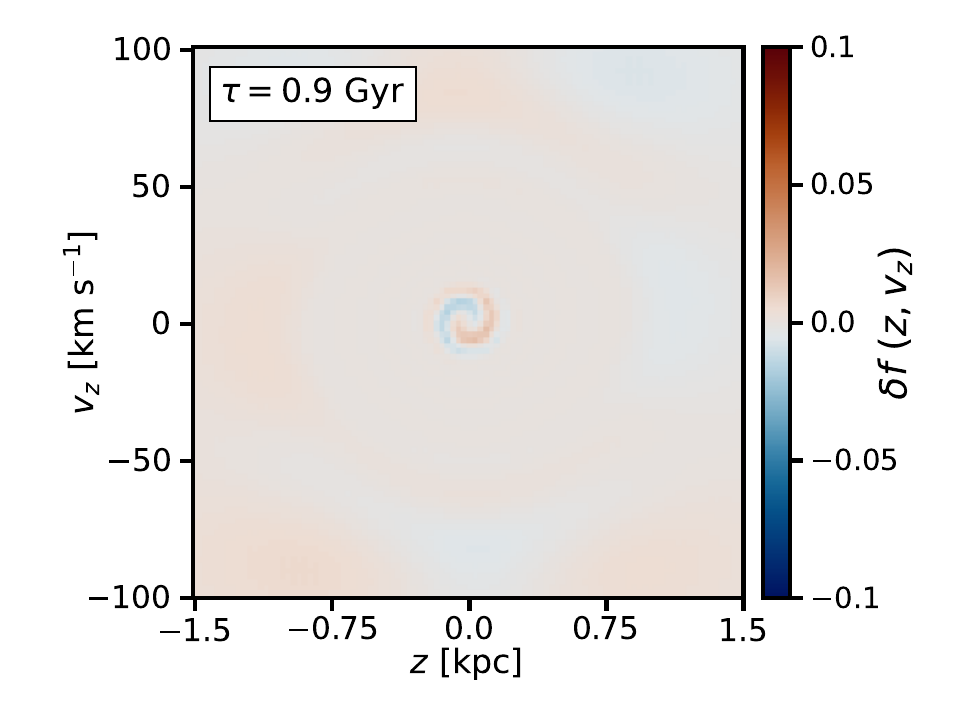}
		\caption{\label{fig:diffusioneffects} The evolution of an impulse perturbation as a function of time using the phenomenological treatment of diffusion discussed in Section \ref{ssec:modeldiffusion}. Each panel shows the perturbed distribution function relative to the equilibrium distribution function, where we use the distribution function given by Equation \eqref{eqn:liwidrowdf}. The time since the application of the impulse perturbation, $\tau$, is shown in each panel. Perturbations decay with time as $\exp\left(-\tau^3 / t_0^3\right)$ with $t_0 = 0.6 \ \rm{Gyr}$ \citep{Tremaine++23} (see also Figure \ref{fig:diffusioncalib}).}
	\end{figure*}
	\begin{figure}
		\centering
		\includegraphics[trim=0.2cm 0.5cm 0.2cm
		0.cm,width=0.48\textwidth]{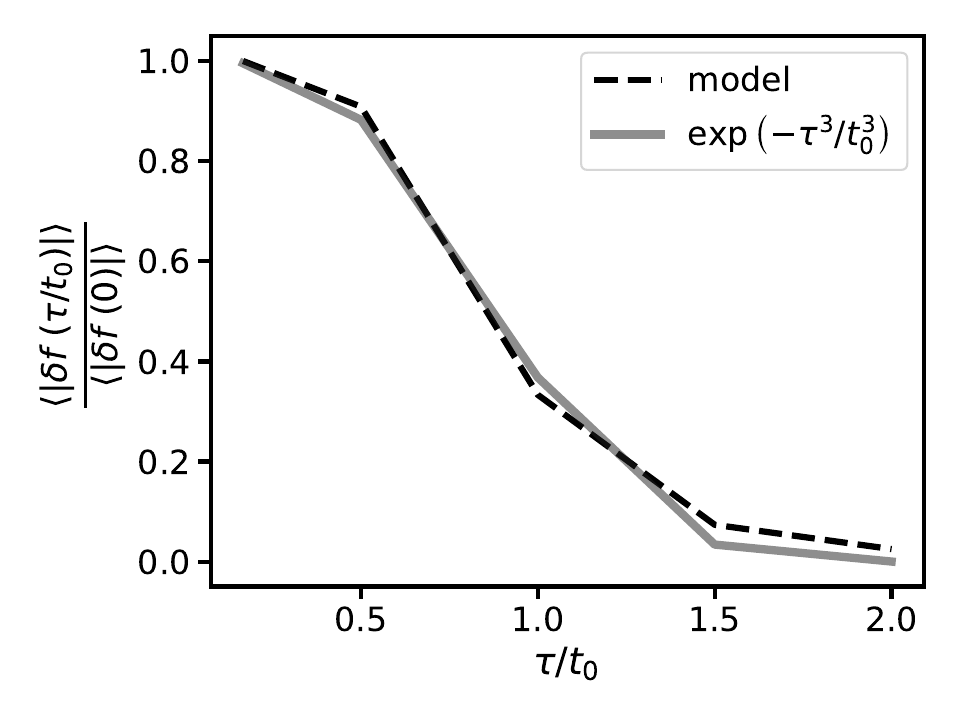}
		\caption{\label{fig:diffusioncalib} The damping of the perturbed distribution functions shown in Figure \ref{fig:diffusioneffects} as a function of $\tau$. The dashed black line shows $\langle | \delta f\left(\tau /t_0\right) | \rangle / \langle | \delta f\left(0\right) | \rangle$, the absolute value of the perturbation to the distribution function averaged across the phase space, as a function of time. We calibrate the model of diffusion to damp the distribution function as $\exp\left(-\tau^3 / t_0^3\right)$, shown by the gray curve.}
	\end{figure}
	
	To fit the parameters $\alpha$ and $\sigma_z$ to Gaia DR3, we use the Gaia RVS subsample with $\varpi / \sigma_\varpi > 5$, and select stars within a projected distance to the Sun on the Galactic midplane of 0.5 kpc. We assume the distance to the Galactic center and the height of the Sun to be $R_\odot = 8.23$ kpc\footnote{The star selection from Gaia DR3 adopted a slightly different distance from the Galactic center than the radius of the solar circle $R_0 = 8.172 \ \rm{kpc}$ where we evaluate the vertical frequency $\nu$. This detail does not impact our main results.} \citep{leung_2022} and $z_\odot = 20.8$ pc \citep{BennettBovy19}, and the solar motion with respect to the Local Standard of Rest to be $v_\odot=(11.1, 12.24, 7.25)\,\mbox{km s}^{-1}$\citep{schoenrich_2010}, with the circular velocity at the solar radius of $220\,\mathrm{km\,s^{-1}}$ as implemented in \texttt{galpy}. We correct the radial velocities according to Equation 5 in \cite{katz_2023}. After fitting the parameters $\alpha$ and $\sigma$ to the vertical velocity dispersion of Gaia DR3, as shown in Figure \ref{fig:vdisfit}, we find $\alpha = 2.34$ and $\sigma_v = 15.20 \ \rm{km} \ \rm{s^{-1}}$. 
	
	\subsection{Diffusion}
	\label{ssec:modeldiffusion}
	As pointed out by \citet{Tremaine++23}, phase-space spirals can result from a continuous series of small perturbations to the vertical dynamics that they modeled as Gaussian noise. The phase spirals that emerge from these small disturbances subsequently diffuse away as a result of gravitational scattering against giant molecular clouds \citep[e.g.,][]{Spitzer++51,Carlberg87,Jenkins++90}. As shown by \citet{Tremaine++23} and \citet{Banik++23}, the timescale for this collisional damping process is $t_0 \sim 0.6 \ \rm{Gyr}$, meaning perturbations older than this should rapidly decay and not contribute to the features of the Gaia snail observed today. 
	
	The model outlined in Section \ref{ssec:modelspirals} does not include a physical treatment of diffusion. However, we can implement a phenomenological treatment of this process, and calibrate the model such that it produces the physical behavior identified by \citet{Tremaine++23}. In particular, we will require that our model reproduce the damping of a perturbation that scales with time as $\exp\left(-\tau^3 / t_0^3\right)$, where $\tau$ is the age of a perturbation. For a passing luminous or dark satellite, we define $\tau$ as the time since the satellite exerted its strongest vertical force on the solar position. 
	
	To construct a model for diffusion, we begin by identifying natural length and velocity scales, $l\left(z, v_z\right)$ and $v\left(z, v_z\right)$, respectively, at each coordinate in the phase space
	\begin{eqnarray}
		l\left(z, v_z\right) &=& \sqrt{J_{\rm{eq}}\left(z, v_z\right)/\Omega_{\rm{eq}}\left(z, v_z\right)}\\
		v\left(z, v_z\right) &=&\sqrt{J_{\rm{eq}}\left(z, v_z\right) \times \Omega_{\rm{eq}}\left(z, v_z\right)}.
	\end{eqnarray}
	We will model diffusion through a series of Gaussian convolutions applied to each $\Delta J_{i}$, the perturbation to the vertical action caused by the $i$th passing satellite. We write the kernels for the convolutions 
	\begin{eqnarray}
		\label{eqn:kernelz}
		k_z\left(z, v_z,\tau/t_0\right) &=&g\left(\tau/t_0\right) l\left(z, v_z\right)\\
		\label{eqn:kernelvz}
		k_{v_z}\left(z, v_z,\tau/t_0\right)  &=& g\left(\tau/t_0\right) v\left(z,v_z\right),
	\end{eqnarray}
	where $g\left(\tau/t_0\right)$ controls how quickly a perturbation decays with time, and then transform $\Delta J_i$ through the operation
	\begin{equation}
		\Delta J_i\left(z, v_z\right) = \int \Delta J_i\left(z^{\prime}, v_z^{\prime}\right)   \mathcal{N}\left(z,z^{\prime}, v_z,v_z^{\prime}\right) dz^{\prime} d v_z^{\prime} 
	\end{equation}
	where
	\begin{equation}
		\mathcal{N}\left(z,z^{\prime},v_z,v_z^{\prime}\right) = \frac{1}{2\pi |\Sigma|^{1/2}} \exp\left(-\frac{1}{2} {x^{T} \Sigma^{-1} x}\right)
	\end{equation}
	with $x = \left(z-z^{\prime}, v_z-v_z^{\prime}\right)$ and $\Sigma = \mathrm{diag}\left[(]k_z^2\left(z, v_z\right), k_{v_z}^2\left(z, v_z\right)\right]$. For a coordinate in phase space centered at $\left(z, v_z\right)$, this is simply a Gaussian convolution with a variance of $k_z$ along the $z$ direction and a variance $k_{v_z}$ along the $v_z$ direction. However, it is not equivalent to a Gaussian convolution of the entire phase-space area due to the spatially varying kernels $k_{z}$ and $k_{v_z}$ given in Equations \eqref{eqn:kernelz} and \eqref{eqn:kernelvz}. 
	
	After some experimentation, we find that a relatively simple normalization of the diffusion kernels
	\begin{equation}
		g\left(\tau/t_0\right) = c_1 \left(\tau / t_0\right)^{c_2}
	\end{equation}
	with $\left(c_1, c_2\right) = \left(0.24, 1.00\right)$ reproduces the $\exp\left(-\tau^3/t_0^3\right)$ damping of perturbations to the distribution functions predicted by \citet{Tremaine++23}. To calibrate this model, we apply impulse perturbations at a series of times spaced between 0  and 1.2 Gyr in the past, calculate $\Delta J$ using the method discussed in Section \ref{ssec:modelspirals}, apply our diffusion model to the $\Delta J$s, and then calculate the distribution function using Equation \eqref{eqn:liwidrowdf}. For each impulse perturbation, we compute the perturbation to the distribution function 
	\begin{equation}
		\label{eqn:dfpert}
		\delta f\left(z,v_z\right) \equiv \frac{f\left(z, v_z\right)}{f_{\rm{eq}}\left(z,v_z\right)}-1
	\end{equation}
	where $f_{\rm{eq}}\left(z,v_z\right)$ represents the distribution function in equilibrium $\left(\Delta J=0\right)$. Using the notation $\delta f\left(\tau/t_0\right)$ to represent the perturbed distribution function $\delta f \left(z, v_z\right)$ that corresponds to each impulse perturbation applied at a time $\tau$, we find values of $c_1$ and $c_2$ such that $\langle |\delta f\left(\tau/t_0\right) | \rangle \propto \exp\left(-\tau^3 / t_0^3\right)$. Here, and throughout this paper, brackets around the distribution function denote an average taken across the entire phase space. Figure \ref{fig:diffusioneffects} shows how the impulse perturbation appears today after application at $\tau = 0.1, 0.3, 0.6, 0.9$ Gyr in the past, subject to the model of diffusion described in the section. In Figure \ref{fig:diffusioncalib}, we show the evolution of $\langle |\delta f\left(\tau/t_0\right)| \rangle$ for the same perturbation. 
	
	\subsection{Perturbing satellites}
	\label{ssec:satellites}
	In this work, we consider a scenario in which luminous and dark Galactic satellites perturb the vertical dynamics near the solar neighborhood. The luminous satellites include the Milky Way's dwarf galaxies with typical masses in the range $10^8 - 10^{10} M_{\odot}$. According to the predictions of CDM, the existence of these luminous satellites implies the presence of a large population of Galactic subhalos \citep{Klypin++99,Moore++99}. Based on the current understanding of galaxy formation, and in particular how the properties of Milky Way satellites depend on their host dark-matter halos, the overwhelming majority of these subhalos are not expected to contain enough luminous material to be detected \citep{Nadler++20}. 
	
\begin{figure*}
		\centering
		\includegraphics[trim=0cm 0cm 0cm
		0cm,width=0.78\textwidth,angle=-90]{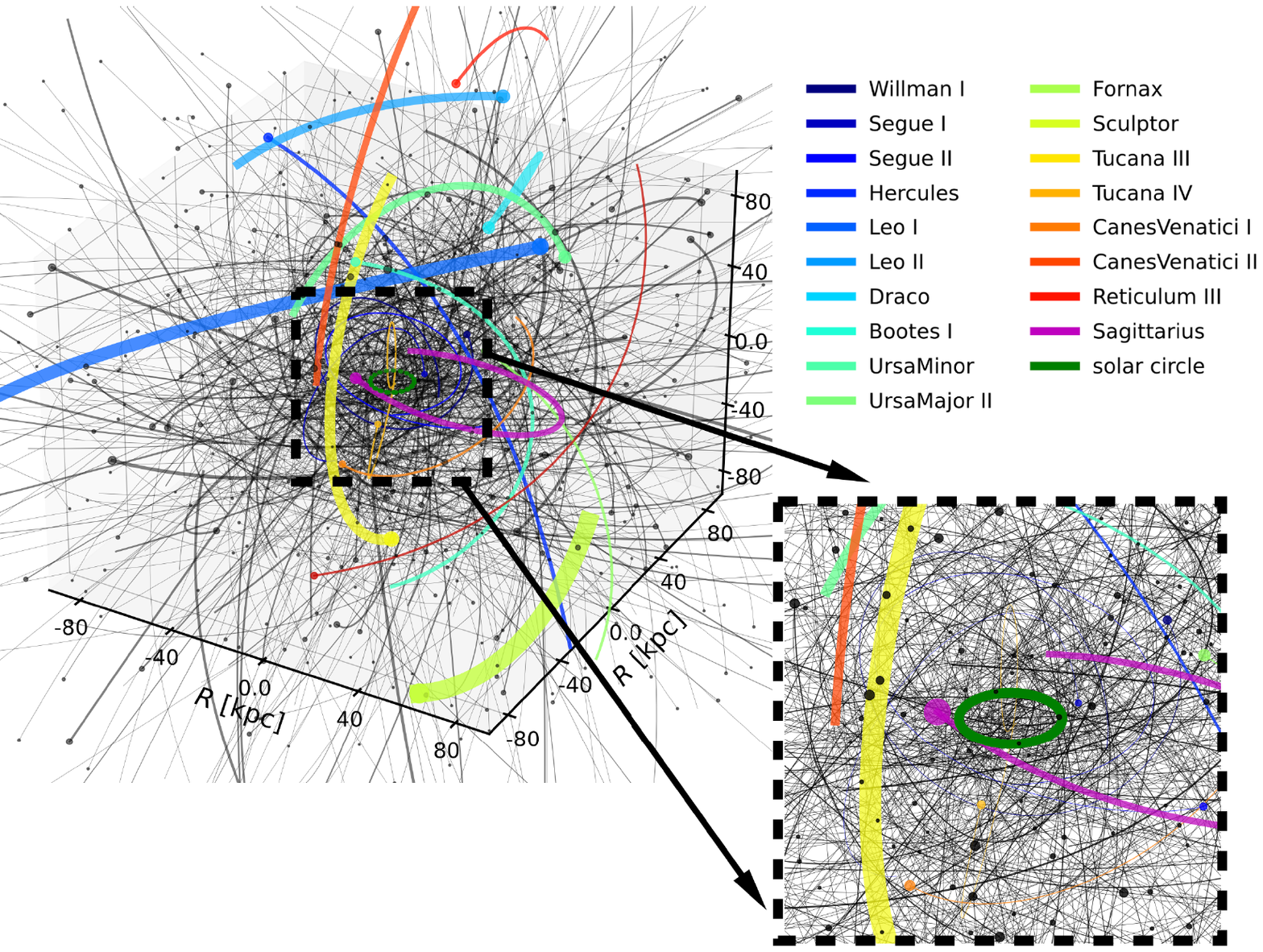}
		\caption{\label{fig:perturberfig} A simulated population of dark subhalos ($10^{6.5} < m_{\rm{sub}} / M_{\odot} < 10^{8}$) and dwarf galaxies, shown as black and colored curves, respectively. Only subhalo orbits passing within  $80$ kpc of the solar position in the past 1.2 Gyr are included. Line width and marker sizes scale with the square root of halo mass. The inset with a dashed-black border highlights in greater detail a 20-by-20 kpc cutout of the solar neighborhood, with the path of the solar system around the Galactic center represented by a dark green circle.}
	\end{figure*}

	The following sub-sections describe how we model the population of dark and luminous satellites. We begin in \ref{sssec:massdef} by discussing the halo mass definition assigned to subhalos, and how uncertainties associated with the mass definition translate to uncertainties in the abundance of subhalos. In \ref{sssec:luminoussatellites}, we describe the treatment of luminous Galactic satellites, including an approximation for their infall masses from abundance matching. In \ref{sssec:darksatellites}, we describe the modeling of dark subhalos, including a probabilistic model for generating their orbits. Figure \ref{fig:perturberfig} provides an illustration of a full population of perturbing satellites created according to the methods outlined in this section. 
	\begin{table}
		\centering
		\caption{Luminous satellite galaxies we include in our simulations. Columns correspond to the peak mass, the maximum vertical force exerted in the last 1.2 Gyr, and the perturbation to the vertical action during this period normalized by $\langle J_{\rm{eq}}\rangle = 11.8 \ \rm{km} \ \rm{s^{-1}} \rm{kpc}$. Forces are given in units of $2 \pi G M_{\odot} \rm{pc^{-2}}$.}
		\label{tab:luminoussats}
		\setlength{\tabcolsep}{4pt}
		\renewcommand{\arraystretch}{1.} 
		\begin{tabular}{lcccr} 
			\hline
			galaxy &  $M_{\rm{peak}}\left[M_{\odot}\right] $ & $\log_{10} |F_{z,\rm{max}}| $ &  $\log_{10} \frac{\Delta J}{\langle J_{\rm{eq}}\rangle}$  \\
			\hline 
			Willman I & $10^{7.7}$ & $-1.7\pm0.3$ & $-2.7\pm0.3$ \\
			Segue I & $10^{7.5}$ & $-1.9\pm0.2$ & $-3.1\pm0.2$ \\
			Segue II & $10^{7.6}$ & $-1.7\pm0.2$ & $-2.8\pm0.2$ \\
			Hercules & $10^{8.3}$ & $-1.9\pm0.1$ & $-3.3\pm$0.2 \\
			Leo I & $10^{9.4}$ & $-1.7\pm0.0$ & $-3.3\pm0.0$ \\
			Leo II & $10^{9.0}$ & $-2.0\pm0.1$ & $-3.1\pm0.1$ \\
			Draco & $10^{8.8}$ & $-1.8\pm0.0$ & $-3.0\pm0.0$ \\
			Bootes I & $10^{8.3}$ & $-1.6\pm0.1$ & $-2.8\pm0.1$ \\
			UrsaMinor & $10^{8.9}$ & $-1.7\pm0.2$ & $-2.8\pm0.2$ \\
			UrsaMajor II & $10^{8.0}$ & $-1.9\pm0.0$ & $-3.2\pm0.0$ \\
			Fornax & $10^{9.7}$ & $-1.3\pm0.0$ & $-2.5\pm0.0$ \\
			Sculptor & $10^{9.2}$ & $-1.2\pm0.0$ & $-2.2\pm0.0$ \\
			Tucana III & $10^{7.7}$ & $-0.7\pm0.3$ & $-1.8\pm0.2$ \\
			Tucana IV & $10^{7.9}$ & $-2.1\pm0.2$ & $-3.3\pm0.2$ \\
			CanesVenatici I & $10^{8.8}$ & $-1.7\pm0.3$ & $-2.8\pm0.3$ \\
			CanesVenatici II & $10^{8.2}$ & $-2.7\pm0.0$ & $-3.9\pm0.0$ \\
			Reticulum III & $10^{7.9}$ & $-2.5\pm0.5$ & $-3.6\pm0.5$ \\
			Sagittarius & $10^{9.7}$ & $-0.6\pm0.2$ & $-1.5\pm0.3$ \\
			\hline		
		\end{tabular}
	\end{table}
	\subsubsection{Halo mass definition}
	\label{sssec:massdef}
	When the Milky Way's host halo absorbs another halo from the field, the accreted halo begins losing mass to tidal stripping. The infall mass, or the mass at accretion, approximately coincides with the peak mass, $M_{\rm{peak}}$. The bound mass refers to the total amount of material that remains gravitationally bound to the the subhalo today. The mass scale most relevant for dynamics, the {\textit{perturbation mass}}, which we will simply refer to as $m$, should be bounded from above by the peak mass and from below by the bound mass. Regarding the lower bound, material removed from a subhalo by tidal stripping could still excite a response in the Galactic disk, even if the stripped material is no longer bound to the subhalo that transported it into the galaxy. Following this reasoning, \citet{Bovy16} showed that tidally-stripped material from a subhalo, as it interacts with a stellar stream, can result in similar observables as a stream interacting with an intact subhalo\footnote{We can imagine a stream of debris from a disrupted halo that transits the solar neighborhood along approximately the same orbit as the halo, leading to a resulting change in the vertical action of nearby stars.}. The upper bound of $M_{\rm{peak}}$ is only approximate for the most massive perturbers because dynamical friction wakes can also trigger a dynamic response in the Galactic disk \citep[e.g.][]{Grand++23}. We do not expect effects associated with dynamical friction to significantly affect dark subhalos, given their low masses at infall.  
	
	As we will discuss in this section, there is a considerable range of theoretical and statistical uncertainty associated with both the total number of in-falling subhalos and their subsequent tidal evolution. We will account for these effects by considering a broad range of tidal mass loss when accounting for the effects of dwarf galaxies. In the case of subhalos, we will explore a broad range of total subhalo abundance to account for uncertainties associated with the total number of objects, and which mass scale (bound versus infall) most strongly correlates with their dynamic effects.
	
	We model all satellites as Navarro-Frenk-White \citep[][hereafter NFW]{Navarro++97} profiles with radial density
	\begin{equation}
		\rho\left(r\right) = \frac{\rho_s}{\left(r / r_s\right)\left(1+r / r_s\right)^2},
	\end{equation}
	where $\rho_s$, $r_s = r_{200} / c$, and $r_{200}$ are the density normalization, scale radius, and virial radius, respectively. We define the viral mass as the mass enclosed inside a sphere of radius $r_{\rm{200}}$ that encloses a mean density $200 \rho_{\rm{crit}}$, where $\rho_{\rm{crit}}$ is the critical density of the Universe. We calculate halo concentrations, $c$, using $c\left(m\right) = 17.5 \left(m/10^8\,M_\odot\right)^{-0.06}$, which closely matches the concentration-mass relation presented by \citet{Diemer++19} and \citet{Johnson++21} at redshift zero. 
	
	\subsubsection{Luminous Galactic satellites}
	\label{sssec:luminoussatellites}
	Table \ref{tab:luminoussats} lists the 18 Milky Way satellite galaxies that we consider in this analysis. We can interpret this subset of perturbers as relatively massive ($m \gtrsim 10^8 M_{\odot}$) subhalos with known orbits. We select this subset of satellites from the Local Volume Database\footnote{https://github.com/apace7/local{\_}volume{\_}database} \citep{Pace++24}\footnote{The remaing sample of dwarfs in the database have either incomplete kinematic information, or do not orbit close enough to be relevant for the Gaia snail.}. To estimate the infall masses of these objects, we use the absolute V-band magnitude to estimate the stellar mass, from which we then predict the peak halo mass using the stellar-mass--halo mass relation for Milky Way satellites presented by \citet{Nadler++20}. The first column of Table \ref{tab:luminoussats} lists the resulting peak mass, $M_{\rm{peak}}$, of each satellite. The assignment of dwarf galaxies masses equal to $M_{\rm{peak}}$ corresponds to an upper limit on the strength of the perturbations they impart to the solar neighborhood. In the case of Sagittarius, the known perturber most relevant for the dynamics of the solar neighborhood due to its close passage in the past Gyr, we calculate $M_{\rm{peak}} \sim 5 \times 10^9 M_{\odot}$, a factor of ten larger than its present-day dynamical mass estimate of $\sim 4 \times 10^8 M_{\odot}$ \citep{VasilievBelokurov20}. 
	
	The second column of Table \ref{tab:luminoussats} lists the maximum vertical force exerted by each satellite, assuming the most likely orbits for these systems and $m = M_{\rm{peak}}$. The third column lists $\langle |J_i| \rangle / \langle J_{\rm{eq}} \rangle$, the strength of the perturbation to the action averaged over the phase space, with $z$ ranging between $-1.5$ and $1.5 \ \rm{kpc}$ and $v_z$ between $-100$ and $100 \ \rm{km} \ \rm{s^{-1}}$. $\langle J_{\rm{eq}}\rangle=11.8 \ \rm{km} \ \rm{s^{-1}} \ \rm{kpc}$ is the mean vertical action in dynamic equilibrium. As suggested by Table \ref{tab:luminoussats}, and as we will show in Section \ref{sec:signatures}, with the exception of Sagittarius, the known luminous satellites of the Milky Way are too distant and low mass to impart strong perturbations to the dynamics of the solar neighborhood. This is the same conclusion drawn by  \citet{Banik++22}, who consider a smaller sample of dwarf galaxies. 
	
	\subsubsection{Dark Galactic satellites}
	\label{sssec:darksatellites}
	We now turn our focus to the remaining population of Galactic satellites that are too small to host a luminous galaxy. These ubiquitous dark objects, which must exist in $\Lambda$CDM, populate the halo mass function from the free-streaming scale, potentially as small as one Earth mass \citep[e.g.,][]{Zheng++24}, up to $\sim 10^8 M_{\odot}$, comparable to the halo mass of the smallest dwarf galaxies estimated from abundance matching and the galaxy-halo connection \citep{Nadler++20}. To examine the dynamic perturbations caused by these structures we require a prescription for generating their orbits and masses.
	
	To build a model for perturbing dark satellites we use the semianalytic model {\tt{galacticus}} \citep{Benson12}. Semianalytic models, such as {\tt{galacticus}}, circumvent the expensive numerical integration of millions or billions of particles, and instead deal directly in terms of derived quantities: halo mass, position, velocity, central density, tidal radius, etc. After calibrating these models with N-body simulations, they can simulate the evolution of galaxies and their dark-matter halos orders of magnitude faster than full N-body simulations while also resolving lower-mass halos without artificial disruption and other numerical systematics\footnote{Of course, semianalytic models will still inherit the limitations of the simulations to which they were calibrated. The version of {\tt{galacticus}} used in this work was calibrated against the {\tt{Caterpillar}} suite of simulations \citep{Griffen++16,Yang++20}}. Despite their computational advantages over N-body simulations, using {\tt{galacticus}} to generate thousands of possbile realizations of subhalo populations, while resolving subhalos down to $5\times 10^5 - 10^6 M_{\odot}$ and tracking their orbits, is still computationally intractable. Our strategy, outlined in this section, will be to use this semianalytic model to calibrate a forward model for subhalo masses and orbits that that we can evaluate on the fly. First, we will calculate $p\left(\boldsymbol{x}, \boldsymbol{v}\right)$, the joint distribution of subhalo positions and velocities today, that {\tt{galacticus}} predicts will bring perturbers within $50  \ \rm{kpc}$ to the Galactic center during the past $2.4$ Gyr. Second, we will calcualte approximately how many subhalos have entered this volume in the past $2.4$ Gyr. 
	
	We perform five simulations of a Milky Way-like galaxy with {\tt{galacticus}} that assume a dark-matter halo mass of $10^{12} M_{\odot}$ at redshift zero. For each halo, we build a merger tree with a minimum halo mass resolution of $5 \times 10^{6} M_{\odot}$ using the algorithm of \cite{Cole++00}, and the branching rate parameters of \cite{Newton++24}. Using the merger history of this tree, we compute concentrations for each halo using the model of \citet{Johnson++21}, and model the density profile of each halo with an NFW profile.
	
	We embed an exponential disk potential within the main branch of the merger tree. This disk has a stellar mass of $10^{10} M_{\odot}$ and a mean specific angular momentum of $10^3\,\mathrm{km\,s}^{-1}\,\mathrm{kpc}$ at redshift zero. Both of these quantities (stellar mass and specific angular momentum) are assumed to scale with time as $\exp(\alpha t)$ where $\alpha=0.1$~Gyr$^{-1}$ and $t=0$ at redshift zero, such that the mass and angular momentum of the disk are smaller in the past. The corresponding radial scale length of the disk (which is assumed to be described by an exponential profile) is found by solving for the radius at which it is rotationally supported in the combined potential of the disk and adiabatically-contracted dark-matter halo, with adiabatic contraction following the model of \cite{Gnedin++04}. The vertical height, $z_\mathrm{s}$, of the disk, which is assumed to follow a $\hbox{sech}^2(z/z_\mathrm{s})$ distribution, is set of $0.137$ times the radial scale length \citep{Kregel++02}. 
	
	{\tt{galacticus}} evolves subhalos from infall until the present day, incorporating effects such as tidal heating and dynamical friction\footnote{For the disk, dynamical friction is computed following the model of \cite{Benson++04}} \citep{Pullen++14,Yang++20,BensonDu22,Du++24}, computed in the time-evolving potential of the combined disk+halo system.  We output the position ${\boldsymbol{x}}$, and velocity ${\boldsymbol{v}}$, of subhalos at redshift zero if two criteria are met: 
	\begin{itemize}
		\item The subhalo must pass within $50 \ \rm{kpc}$ of the Galactic center in the past $2.4 \ \rm{Gyr}$. We apply this selection criterion to present-day subhalos that were also subhalos at $t = -2.4 \  \rm{Gyr}$, and to present-day subhalos that were infalling at $t = -2.4 \ \rm{Gyr}$, meaning they were not formally subhalos at $t = -2.4 \ \rm{Gyr}$ but have since accreted onto the Milky Way's host halo and passed near the Galactic center. 
		\item A subhalo must have a bound mass above $5 \times 10^6 M_{\odot}$ at $t = 0 \ \rm{Gyr}$. Extremely tidally disrupted subhalos are therefore removed from the sample when their bound mass drops below the resolution of the merger tree. Our estimation of $\boldsymbol{x}$ and $\boldsymbol{v}$ therefore excludes regions of parameter space that result in severe tidal disruption. For the subhalos that pass within $50 \ \rm{kpc}$ of the Galactic center and survive, {\tt{galacticus}} predicts a median mass loss for subhalos $m_{\rm{bound}} / m_{\rm{infall}} = 0.15_{-0.04}^{+0.06}$, where the uncertainties reflect scatter around the median value.
	\end{itemize}
	The set of subhalos that meet both of these selection criteria constitute the objects most likely to impart a significant perturbation to the solar neighborhood. Our goal now is to generate new realizations of subhalo populations with unique orbital trajectories and subhalo masses. We will assume that orbits of subhalos are independent of their masses\footnote{Dynamical friction can correlate halo masses and orbits, but this process is negligible for the low-mass objects of interest.} and treat these quantities independently. 
	
	We obtain a statistical representation of the orbital trajectories of dark subhalos by applying a Gaussian kernel density estimator (KDE) to the samples from ${\bf{x}}$ and ${\bf{v}}$, the position and velocity of subhalos that meet the two criteria outlined above. The KDE forms an approximation of $p\left(\boldsymbol{x}, \boldsymbol{v}\right)$, the joint distribution of subhalo position and velocity at $t = 0$ (we select the orbital properties at $t=0$ because Equation \ref{eqn:dji} dictates we integrate their orbits backwards in time). Assuming each subhalo's orbit is independent, we can then generate a new realization of subhalo orbits by resampling from the KDE. 
	
	The total number of subhalos generated through this approach is determined by a parameter $\eta$, which sets the amplitude of the perturber mass function
	\begin{equation}
		\label{eqn:dndm}
		\frac{dN}{d m} = \frac{\eta}{m_0}  \left(\frac{m} {m_0}\right)^{-\alpha}
	\end{equation}
	where $\alpha = 1.9$ \citep{Springel++08}, we choose $m_0 = 10^8 M_{\odot}$, and here $m$ refers to the perturbation mass of a subhalo, as discussed in Section \ref{sssec:massdef}. 
	
	Our strategy for determining $\eta$ will be to generate subhalo masses from Equation \ref{eqn:dndm} in the range $10^7-10^8 M_{\odot}$, integrate their orbits in the {\tt{MWPotential2014}} implemented in {\tt{galpy}}, and then calculate the number of subhalos that pass within $50 \ \rm{kpc}$ of the solar position in the past 2.4 Gyr, $N$, relative to the {\tt{galacticus}} prediction, $N_{\rm{galac}}$. The number of subhalos predicted by {\tt{galacticus}} will not be fully converged near the minimum mass resolution of the merger trees $5\times10^6 M_{\odot}$, so we make the comparison between $N$ and $N_{\rm{galac}}$ above this threshold, from $10^7 M_{\odot}-10^{8} M_{\odot}$. 
	
	{\tt{galacticus}} predicts that $N_{\rm{galac}} = 249 \pm 65$ subhalos with bound masses in the range $10^7 - 10^8 M_{\odot}$ pass within $50 \ \rm{kpc}$ of the Galactic center in the past 2.4 Gyr. The uncertainty in the number of subhalos reflects the halo-to-halo variation between the five {\tt{galacticus}} runs. If we instead assume the mass scale most relevant for the dynamical perturbation is the peak mass, then we increase the number of perturbers to $N_{\rm{galac}} = 1743 \pm 455$. We expect another factor of two uncertainty from the total mass of the Milky Way's dark-matter halo \citep{Wang++20}.  
	
	With our approximation of $p\left(\boldsymbol{x}, \boldsymbol{v}\right)$, we find $N \approx 324 \left(\eta/1000\right)$. For reference, \citet{Feldmann++15} estimated 500 subhalos more massive than $10^7 M_{\odot}$ have passed within 25 kpc of the solar neighborhood in the past 2 Gyr, corresponding to $\eta \gtrsim 1500$. The population of dark satellites shown in Figure \ref{fig:perturberfig} corresponds to $\eta = 1100$. The uncertainties in this parameter are derived primarily from ambiguity in the appropriate halo mass definition and tidal stripping, as discussed in Section \ref{sssec:massdef}. The upper range of $\eta$ corresponds to assigning subhalos $m = M_{\rm{peak}}$ and assuming a more-massive Milky Way host halo. For the lower bound, while we expect the Milky Way's disk to accelerate the tidal disruption of halos \citep[e.g.][]{Garrison-Kimmel++17,Wang++24}, which for the {\tt{galacticus}} simulations means the subhalo bound mass drops below the resolution limit, the tidally stripped material from the halo could still excite a response in the disk if it remains in close proximity to the subhalo from which it was removed. We therefore expect tidal stripping to impact the detectable signal from subhalos only if the tidally stripped material has completely disassociated from the subhalo, which could occur if the subhalo crosses the Galactic disk long before its interaction with the solar neighborhood. However, we find that only $\sim 10\% $ of perturbers execute these kinds of orbits, as predicted by our model for $p\left(\boldsymbol{x}, \boldsymbol{v}\right)$. To further investigate uncertainties associated with the model of the Galactic disk, we have run {\tt{galacticus}} with a more-massive stellar disk, assuming $m_{\rm{disk}}=6.8 \times10^{10} M_{\odot}$ at redshift zero. This more-massive disk depletes the abundance of subhalos predicted by {\tt{galacticus}} by a further $15 \%$, relative to assuming $M_{\rm{disk}} = 10^{10} M_{\odot}$ at redshift zero. Given the intrinsic scatter between the {\tt{galacticus}} runs, the total mass of the Milky Way's host halo, and the factor of 7 difference between the bound mass and peak mass definitions, we will explore a range of $\eta$ that spans a factor of 24: $\eta \sim 500 - 12000$.
	
	We note that the orbits of subhalos in the {\tt{galacticus}} simulations and those we use in our model will differ due to the assumptions regarding the gravitational potential of the dark-matter halo, disk, and bulge. To account for systematics associated with the different potentials we track the orbits that come within $R = 50 \ \rm{kpc}$ of the Galactic center, recording subhalo trajectories that enter a significantly larger volume than what we expect will be relevant for the Gaia snail. These complications aside, the orbital properties of subhalos generated through our approach appear broadly consistent with those of known Milky Way satellites, as discussed in the next section. \\
	
	\section{Dark perturbations to the Gaia snail}
	\label{sec:signatures}
	In this section, we use the model described in Sections \ref{ssec:modelspirals} and \ref{ssec:modeldiffusion}, and \ref{ssec:satellites} to calculate distribution functions for stars in the solar neighborhood subject to perturbations by luminous and dark Galactic satellites. We begin in Section \ref{ssec:impactstats} by examining the distribution of vertical forces and perturbations to the vertical action caused by dark and luminous satellites. In Section \ref{ssec:impactobs}, we calculate the perturbed phase-space distributions and summary statistics that encode the observational signatures of disequilibrium. Section \ref{ssec:dfstats} compares the predictions of our model with observations from Gaia. 
	\begin{figure*}
		\includegraphics[trim=0cm 0.7cm 0cm
		0.5cm,width=0.95\textwidth]{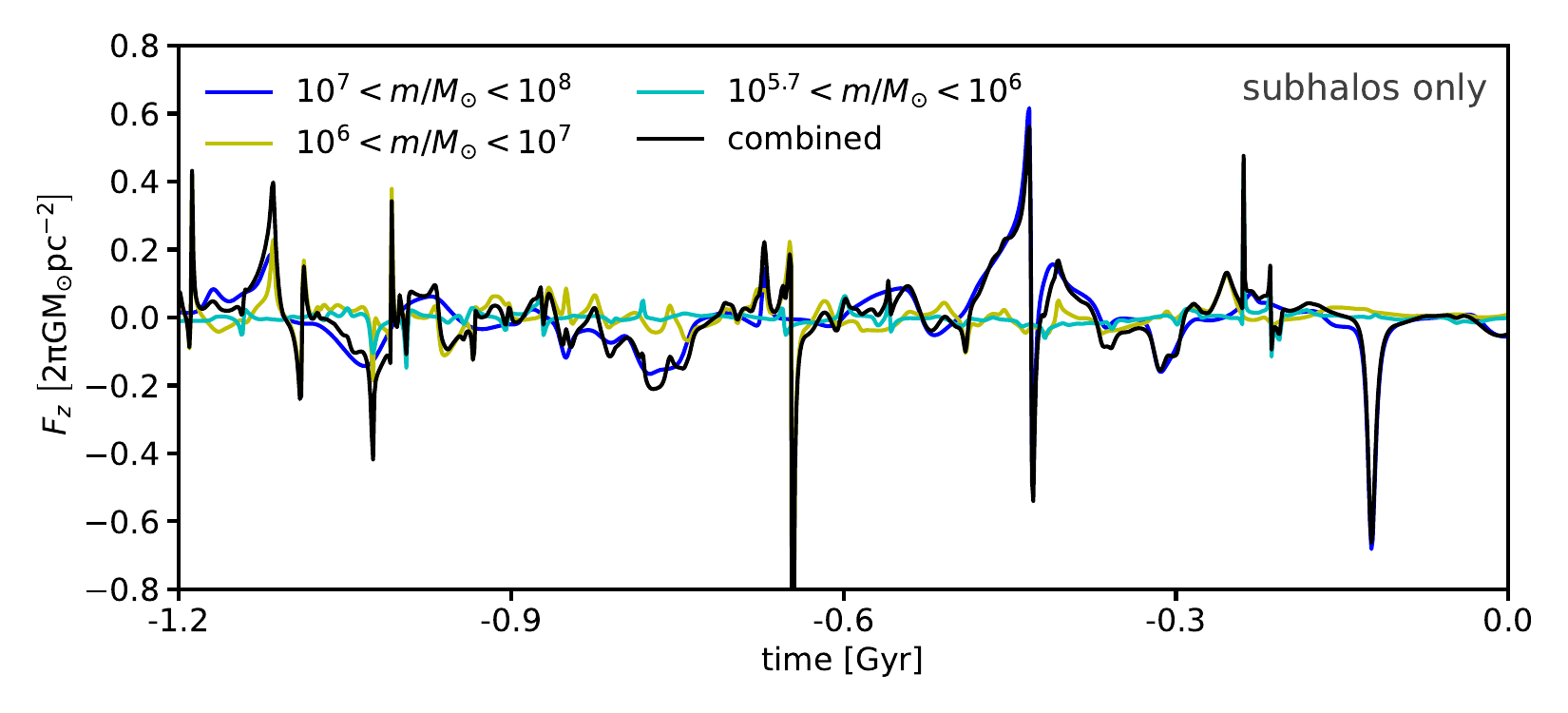}
		\includegraphics[trim=0cm 0.7cm 0cm
		0cm,width=0.95\textwidth]{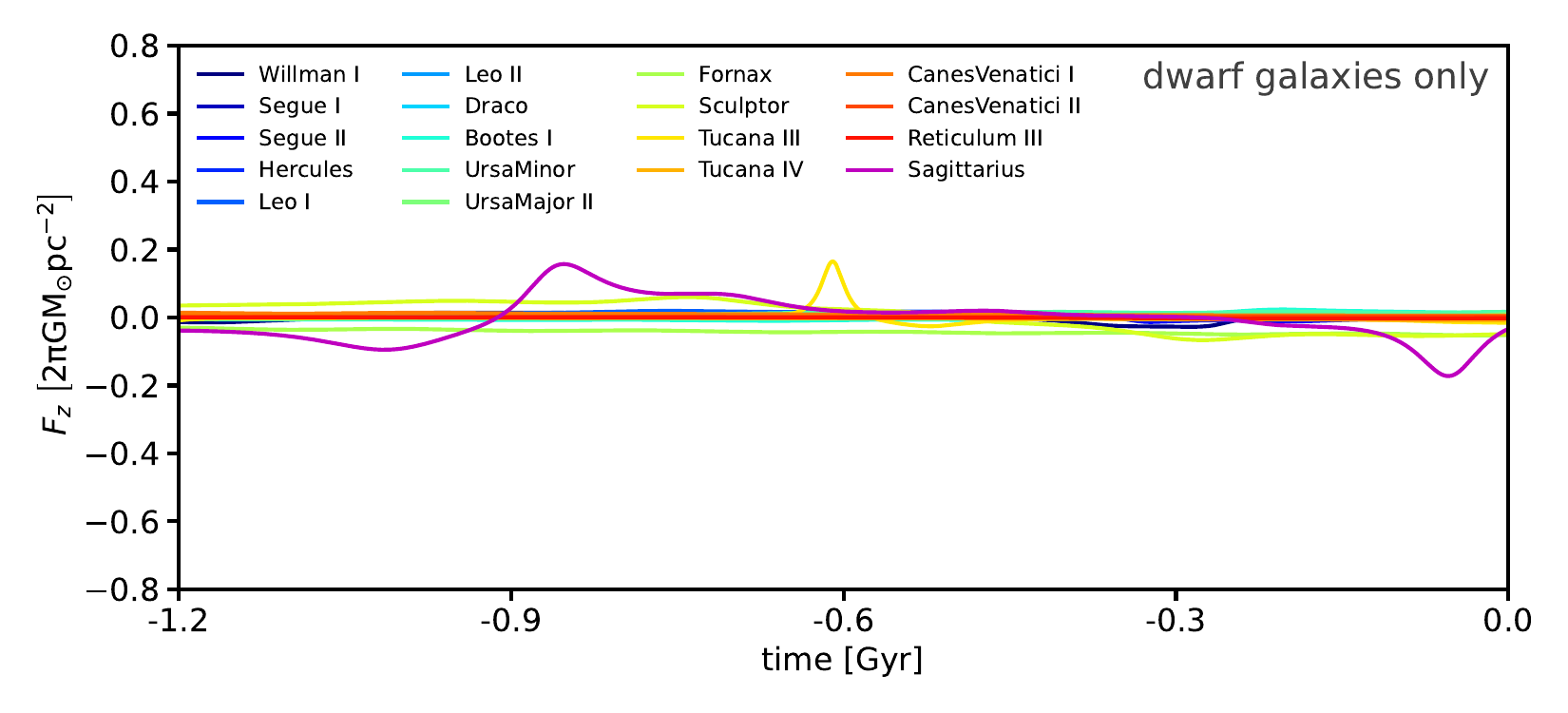}
		\includegraphics[trim=0cm 0.7cm 0cm
		0cm,width=0.95\textwidth]{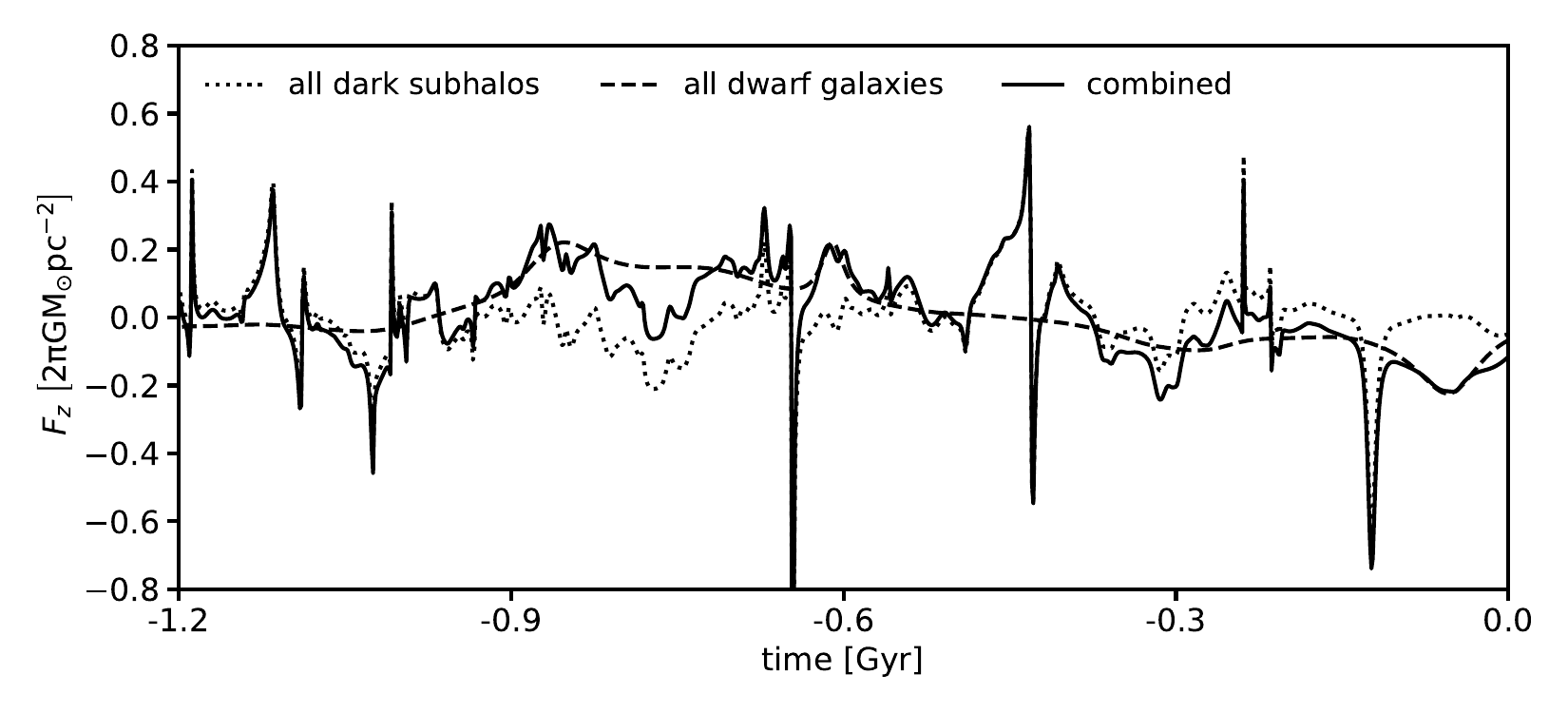}
		\caption{\label{fig:forcefig} The force exerted as a function of time by a realization of dark subhalos (top), dwarf galaxies (middle), and the combination of both populations of satellites (bottom). Colors in the top panel correspond to different subhalo mass ranges, while colors in the middle panel differentiate the dwarf galaxies listed in Table \ref{tab:luminoussats}. }
	\end{figure*}
	\begin{figure*}
		\includegraphics[trim=0cm 0.5cm 0cm
		0cm,width=0.48\textwidth]{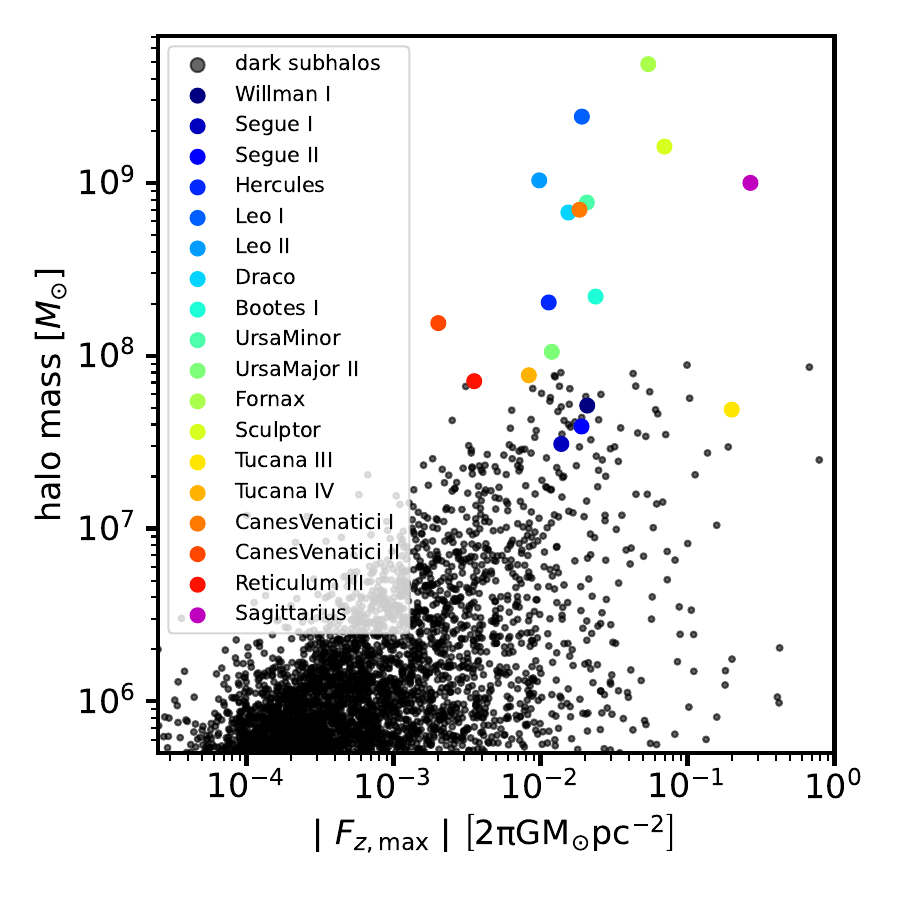}
		\includegraphics[trim=0cm 0.5cm 0cm
		0cm,width=0.48\textwidth]{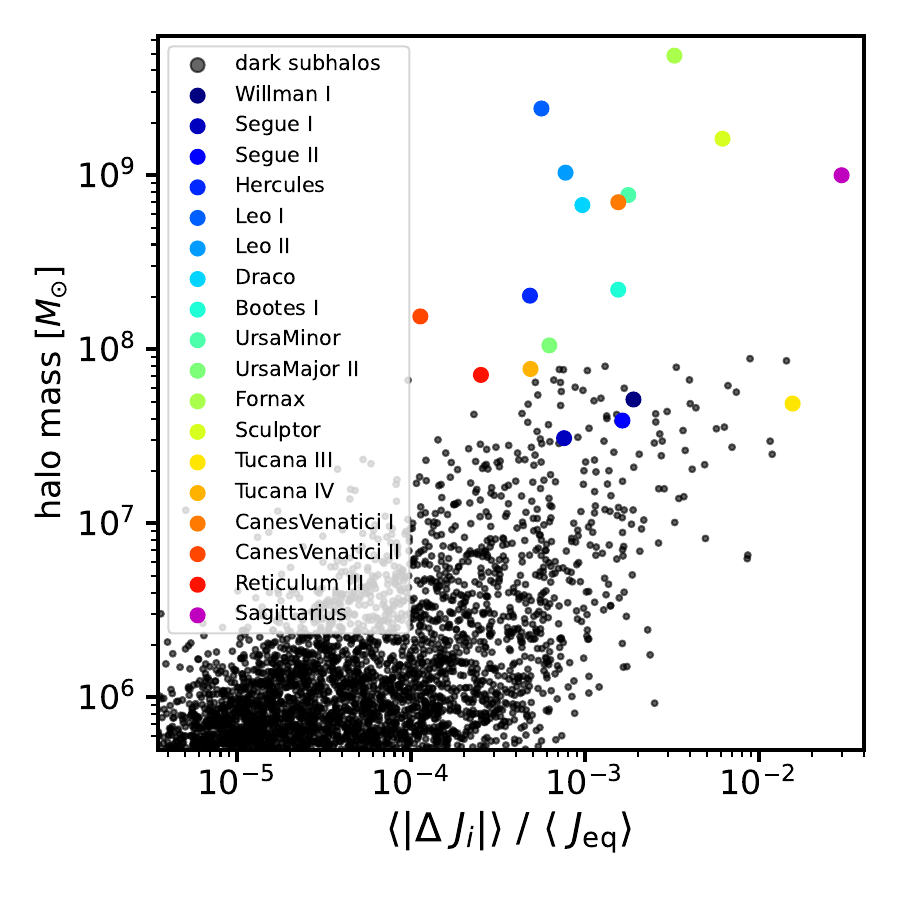}
		\caption{\label{fig:fmaxvshalomass} {\bf{Left:}} The joint distribution of the maximum vertical force exerted by a dark or luminous satellite in the past 2.4 Gyr and halo mass. Some low-mass satellites exert a maximum vertical force comparable to that of luminous satellites if their orbits bring them in close proximity to the solar circle. {\bf{Right:}} The joint distribution of halo mass and the average perturbations to the vertical action, $|\langle \Delta J_{i}\rangle |$, where the brackets denote averaging over the entire phase space, and we normalize by the mean vertical action in the solar neighborhood in dynamic equilibrium, $\langle J_{\rm{eq}} \rangle \sim 11.8 \ \rm{kpc} \ \rm{km} \ \rm{s^{-1}} $. The perturbation to the vertical action is calculated by integrating satellite orbits following the procedure outlined in Section \ref{ssec:modelspirals}. }
	\end{figure*}
	\begin{figure*}
		\includegraphics[trim=0cm 0.5cm 0cm
		0cm,width=0.48\textwidth]{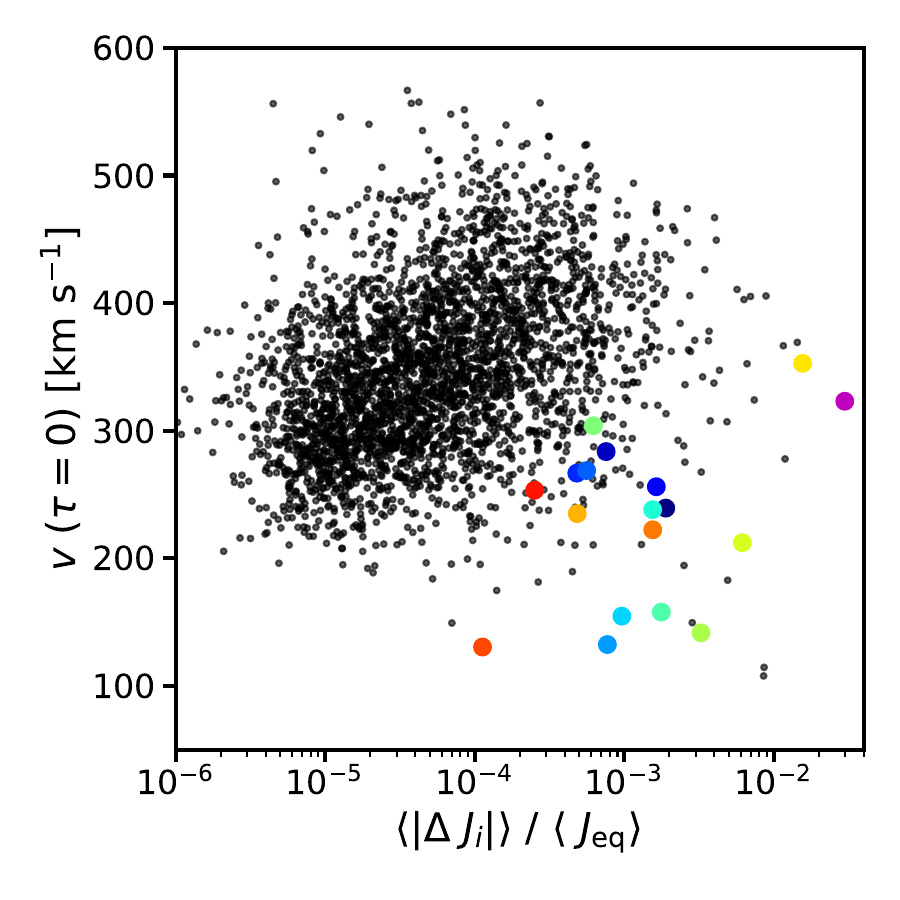}
		\includegraphics[trim=0cm 0.5cm 0cm
		0cm,width=0.48\textwidth]{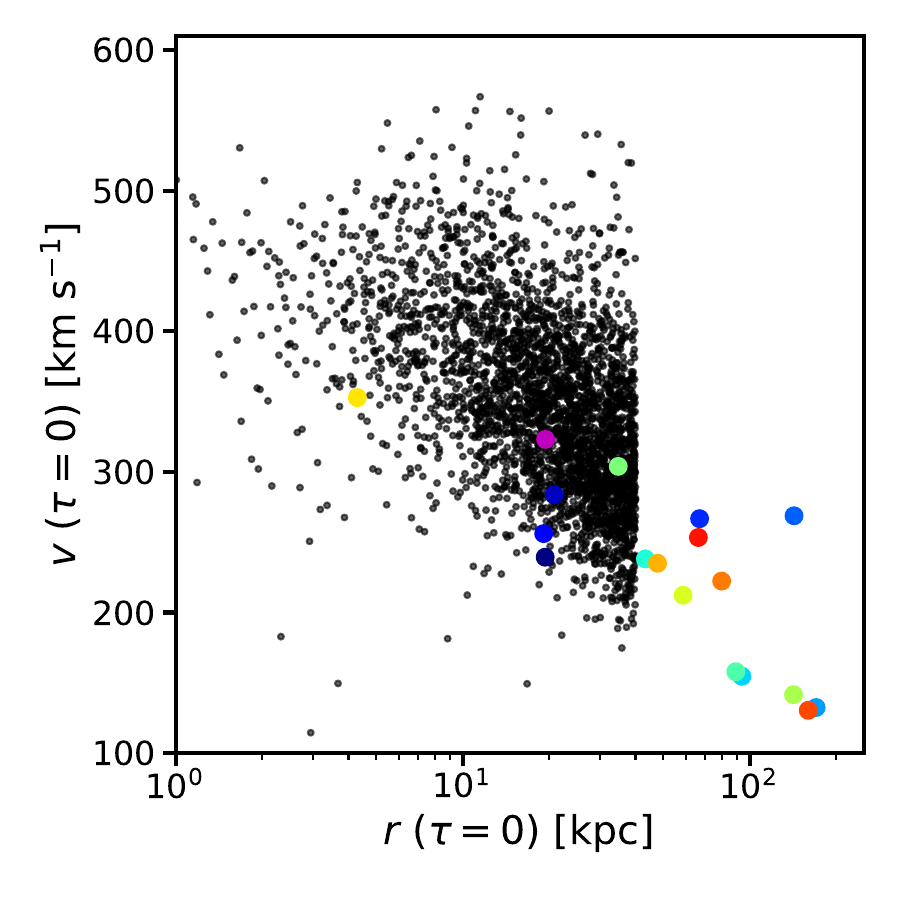}
		\caption{\label{fig:vimpact} {\bf{Left:}} The joint distribution of the average perturbation to the vertical action by a satellite normalized by the mean vertical action in equilibrium (the same quantity as in the right-hand panel of Figure \ref{fig:fmaxvshalomass}) and the velocity of the satellite at the impact time $\tau = 0$. Impact time is defined as the time when a satellite exerts its strongest vertical force on the solar neighborhood in the past 1.2 Gyr. {\bf{Right:}} The joint distribution of the velocity at impact time and the distance from the solar neighborhood at the impact time. We impose a hard cut at $40 $ kpc when rendering subhalos, so no objects have a minimum distance from the solar neighborhood greater than this value. The color scheme is the same as in Figure \ref{fig:fmaxvshalomass}. }
	\end{figure*}
	
	\subsection{Subhalo impact statistics}
	\label{ssec:impactstats}
	Figure \ref{fig:forcefig} shows the vertical force exerted by a single realization of dark subhalos and the luminous satellites listed in Table \ref{tab:luminoussats} as a function of time. The abundance of dark subhalos in this realization corresponds to $\eta = 1500$. The top row illustrates the differential contribution to the vertical force as a function of subhalo mass, excluding the forces exerted by dwarf galaxies. The middle panel shows the force exerted by luminous satellites, assuming perturbation masses equal to the peak masses listed in Table \ref{tab:luminoussats}. The only exception is Sagittarius, to which we have assigned $m = 10^9 M_{\odot}$. The bottom panel of Figure \ref{fig:forcefig} shows the combined effects of subhalos and dwarf galaxies. 
	
	Figures \ref{fig:fmaxvshalomass} and \ref{fig:vimpact} show the statistics of subhalo impacts to the solar neighborhood, including perturbing forces, changes to the vertical action, and orbital properties. These figures include the same population of subhalos shown in Figure \ref{fig:forcefig}.  Figure \ref{fig:fmaxvshalomass} shows the joint distribution of subhalo mass and the maximum vertical force exerted by each object in the past 1.2 Gyr (left), and the joint distribution of halo mass and $\langle |\Delta J_i| \rangle$, the average magnitude of the perturbation to the vertical action, normalized by $\langle J_{\rm{eq}}\rangle$. The left panel of Figure \ref{fig:vimpact} shows the joint distribution of the speed at the impact time $\tau = 0$ and $\langle |\Delta J_i| \rangle$. The right panel shows the joint distribution of $v\left(\tau=0\right)$ and $r\left(\tau=0\right)$, the speed and distance of a perturber when it exerts its strongest vertical force on the solar position. At the impact time $\tau=0$, subhalos frequently come within $10 \ \rm{kpc}$ of the solar position at speeds exceeding $300 \ \rm{km} \ \rm{s^{-1}}$. This results in the series of strong ($|F_{\rm{max}}| > 0.1 \ 2 \pi \rm{G} M_{\odot} \rm{pc^{-2}}$), short-duration impulses visible in Figure \ref{fig:forcefig}, and contrasts with the forces exerted by dwarf galaxies, which tend to have speeds slower than $300 \ \rm{km} \ \rm{s^{-1}}$ due to their wider orbits. Our model prediction for the speeds of dark subhalos passing near the solar neighborhood is consistent with the analytic predictions by \citet{Feldmann++15}. We also note that the speeds of dark satellites predicted by our model coincide with the velocity scale $\sim 2$ to $3\, v_{\rm{cir}} \sim 400$ to $600 \ \rm{km} \ \rm{s^{-1}}$, where $v_{\rm{circ}}$ is the circular velocity of the solar neighborhood. \citet{Banik++23} identified these interaction velocities as those that tend to excite the strongest responses in the Galactic disk. 
	
	As illustrated by Figure \ref{fig:fmaxvshalomass}, a randomly selected dwarf galaxy will exert a significantly stronger maximum vertical force, and be therefore be associated with a larger $\langle |\Delta J| \rangle$, than a randomly selected subhalo\footnote{The uncertainties in the maximum vertical forces and $\langle | \Delta J_i | \rangle$ for luminous satellites are typically the level of a few percent. The exception is Sagittarius, whose orbital uncertainties lead to a $\sim 50\%$ uncertainty in the maximum vertical force. This does not change our conclusions regarding the relative importance of luminous and dark satellites.}. However, the large number of dark subhalos makes it more likely that an object exerting a strong vertical force on the solar neighborhood belongs to this population. For the particular realization of subhalos shown in Figure \ref{fig:fmaxvshalomass}, more than twice as many subhalos impart perturbations stronger than $\langle |\Delta J_i| \rangle / \langle J_{\rm{eq}}\rangle = 5\times 10^{-3}$ than do dwarf galaxies. Our comparison with dwarf galaxies assumes they have present-day masses $m=M_{\rm{peak}}$, with $M_{\rm{peak}}$ given in Table \ref{tab:luminoussats}. If luminous satellites lose a significant fraction of their mass, as expected for halos orbiting in a central potential, they impart weaker perturbations than those shown in Figures \ref{fig:fmaxvshalomass} and \ref{fig:vimpact} and become less relevant for understanding the dynamics of Gaia snail.
	\begin{figure*}
		\includegraphics[trim=1.cm 0.cm 1cm
		2cm,width=0.48\textwidth]{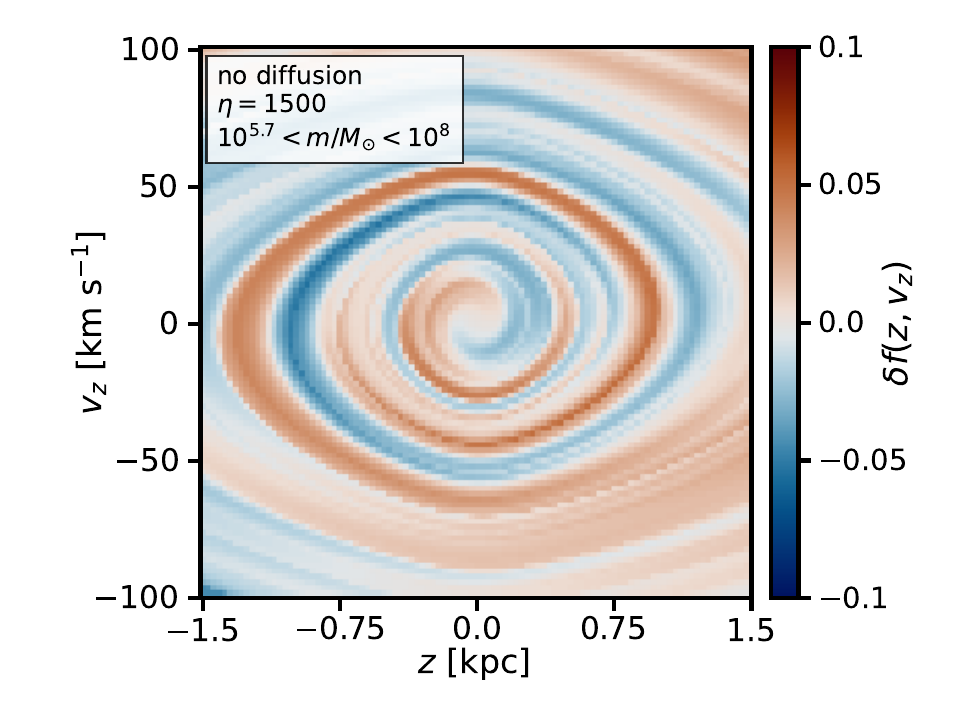}
		\includegraphics[trim=1.cm 0.cm 1cm
		2cm,width=0.48\textwidth]{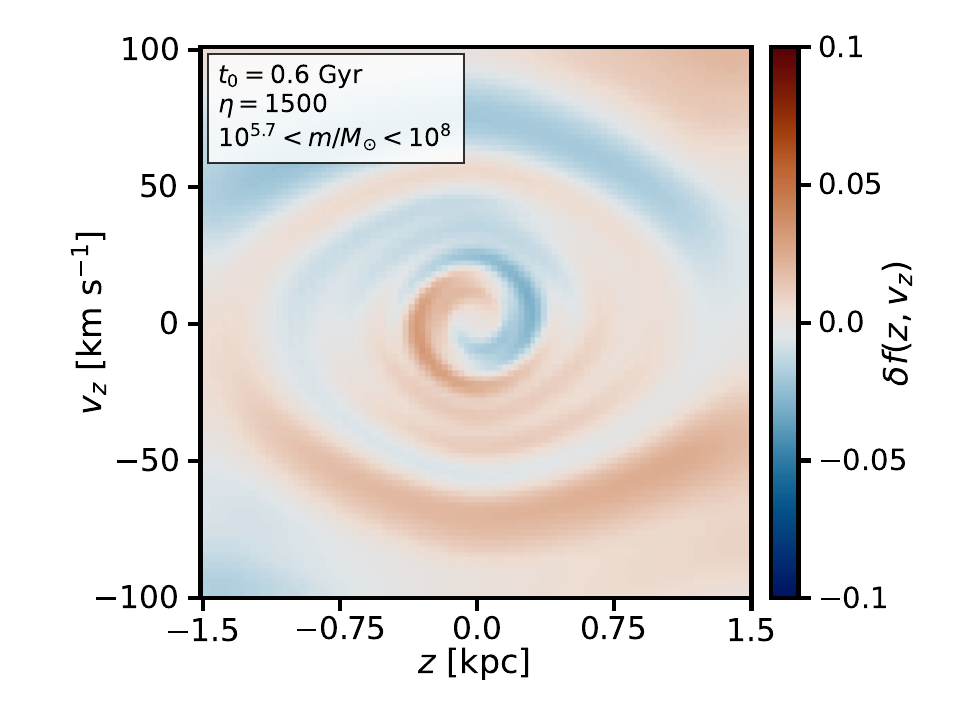}
		\includegraphics[trim=0.cm 0.cm 0.5cm
		0.25cm,width=0.48\textwidth]{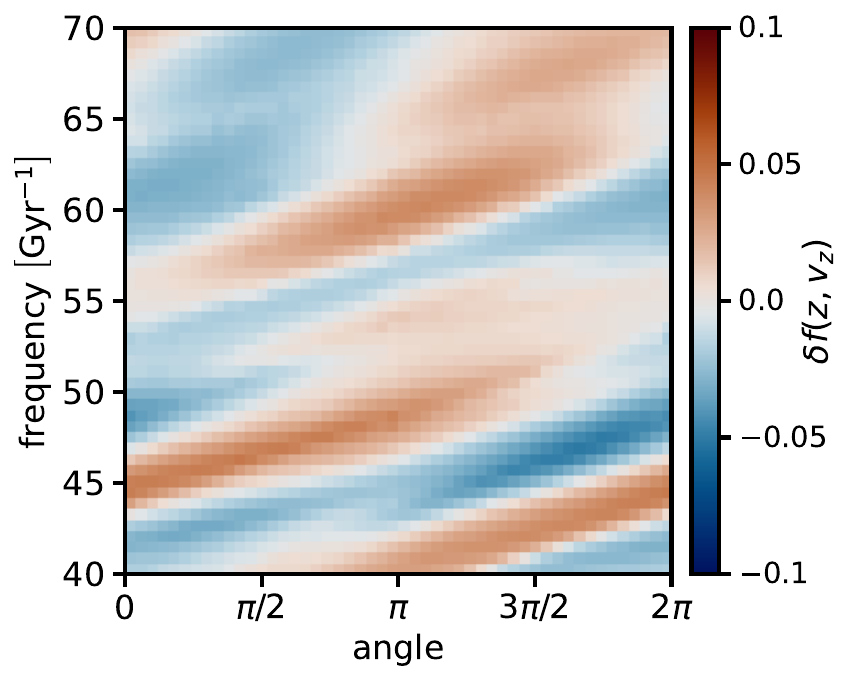}
		\includegraphics[trim=0.cm 0.cm 0.5cm
		0.25cm,width=0.48\textwidth]{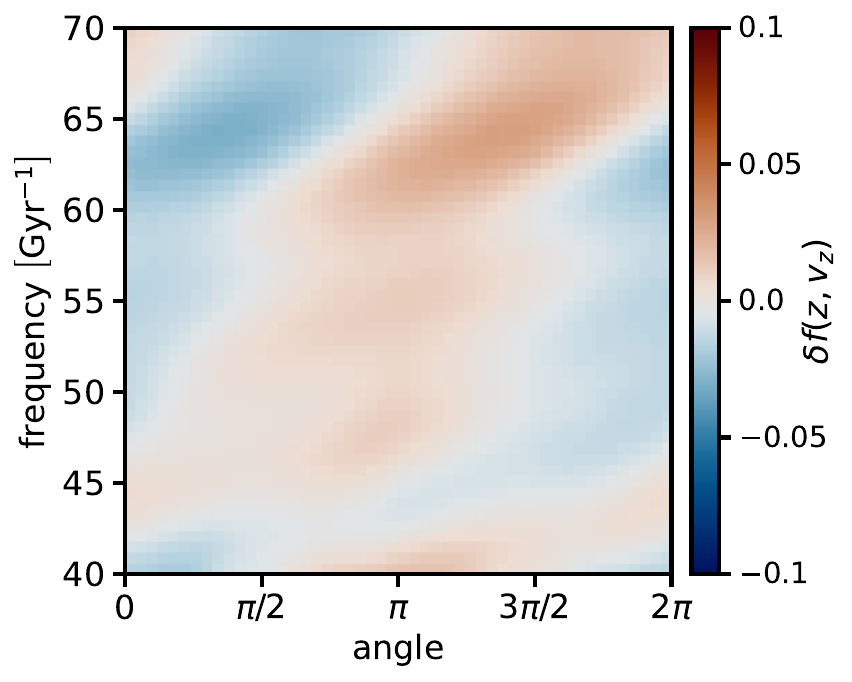}
		\includegraphics[trim=0.25cm 0.5cm 0.5cm
		0.cm,width=0.48\textwidth]{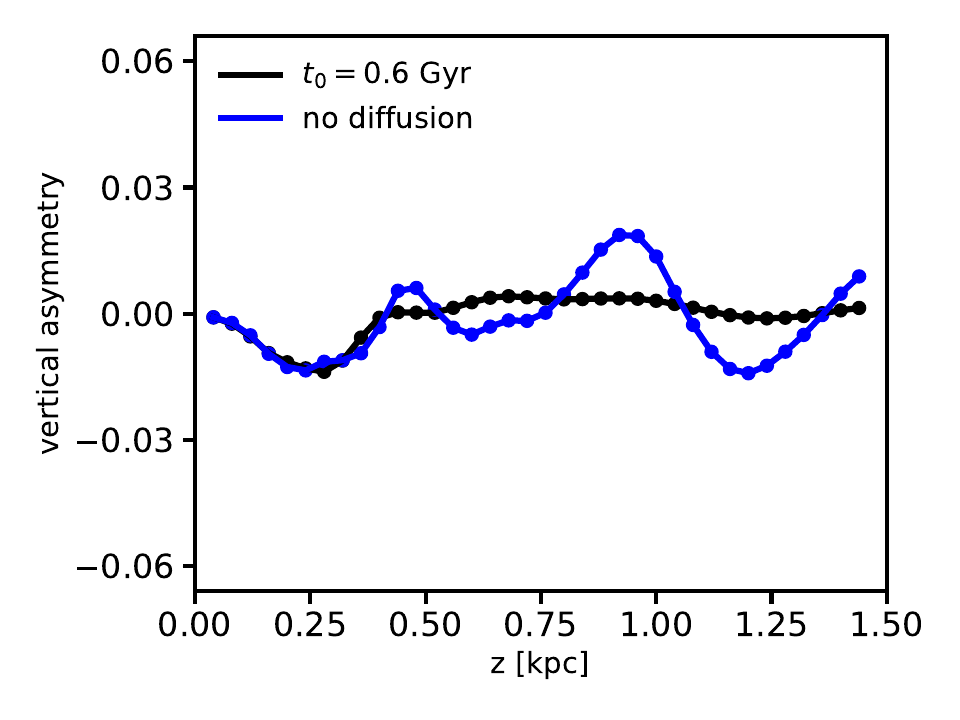}
		\includegraphics[trim=0.25cm 0.5cm 0.5cm
		0.cm,width=0.48\textwidth]{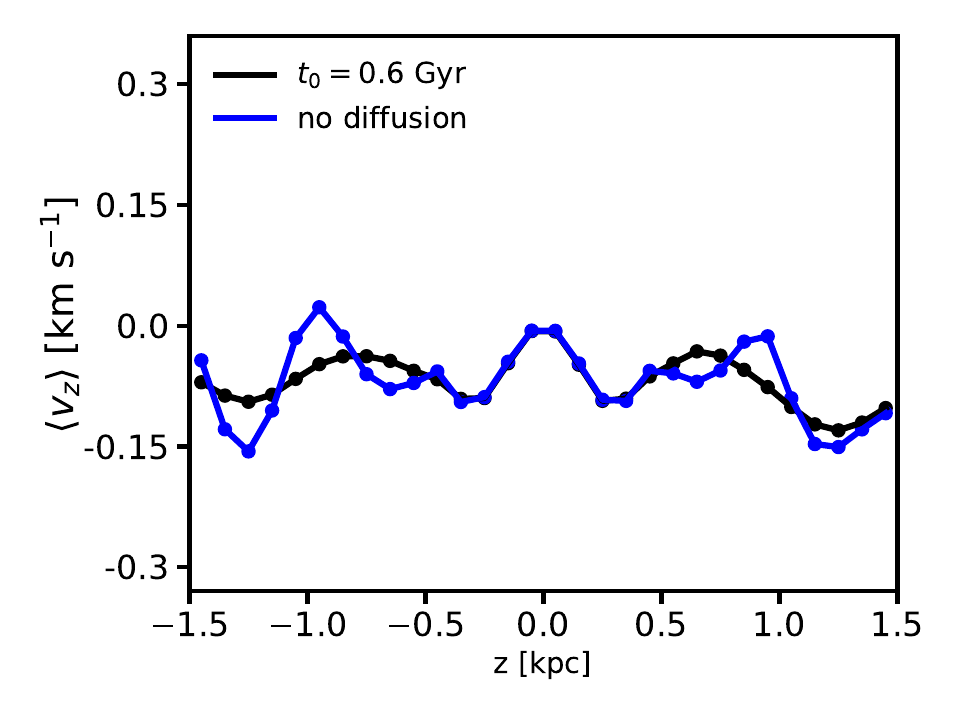}
		\caption{\label{fig:dfexample1992} The distribution function perturbed by the same population of dark subhalos ($10^{5.7} < m / M_{\odot}< 10^8$ and $\eta = 1500$) whose vertical force and orbital properties are shown in Figures \ref{fig:forcefig}-\ref{fig:vimpact}. To highlight the effects of subhalos, effects from luminous satellites are not included. The top row shows the distribution function in $\left(z, v_z\right)$ coordinates, and in the second row we use $\left(\Omega, \theta\right)$ coordinates. The bottom row shows the vertical asymmetry (left) and mean vertical velocity (right). The left column (blue curves) show the perturbation without accounting for diffusion, and the right column (black curves) use the model for diffusion described in Section \ref{ssec:modeldiffusion}.}
	\end{figure*}
	\begin{figure*}
		\includegraphics[trim=1.cm 0.cm 1cm
		2cm,width=0.48\textwidth]{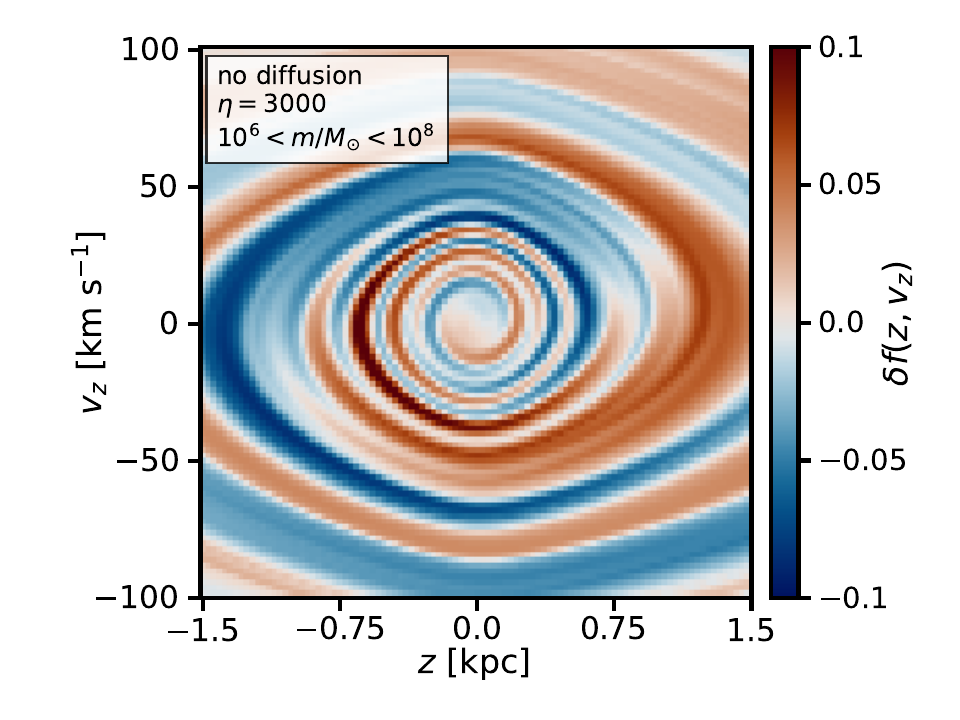}
		\includegraphics[trim=1.cm 0.cm 1cm
		2cm,width=0.48\textwidth]{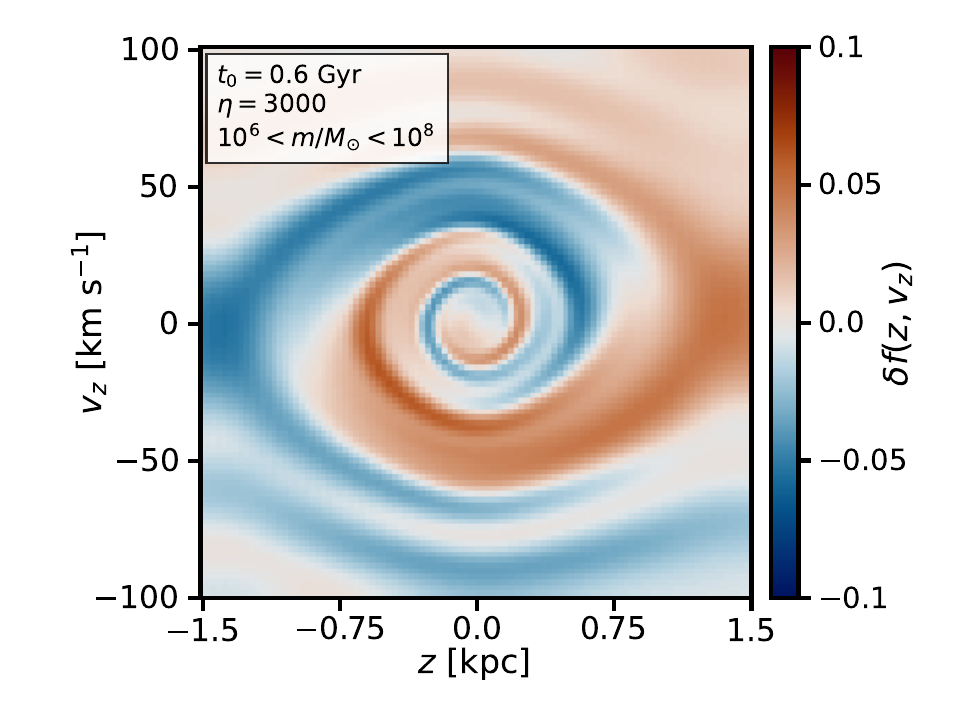}
		\includegraphics[trim=0.cm 0.cm 0.5cm
		0.25cm,width=0.48\textwidth]{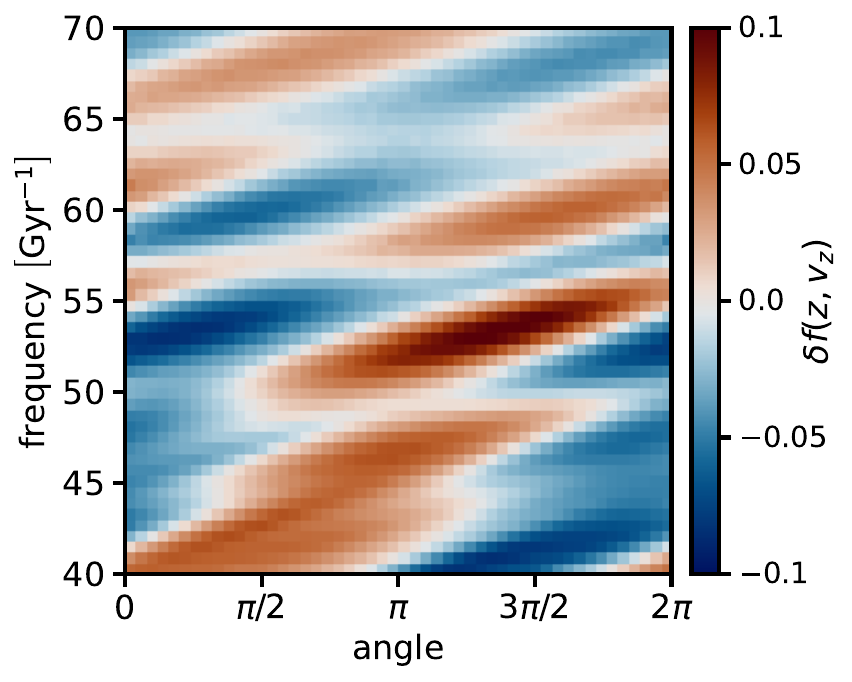}
		\includegraphics[trim=0.cm 0.cm 0.5cm
		0.25cm,width=0.48\textwidth]{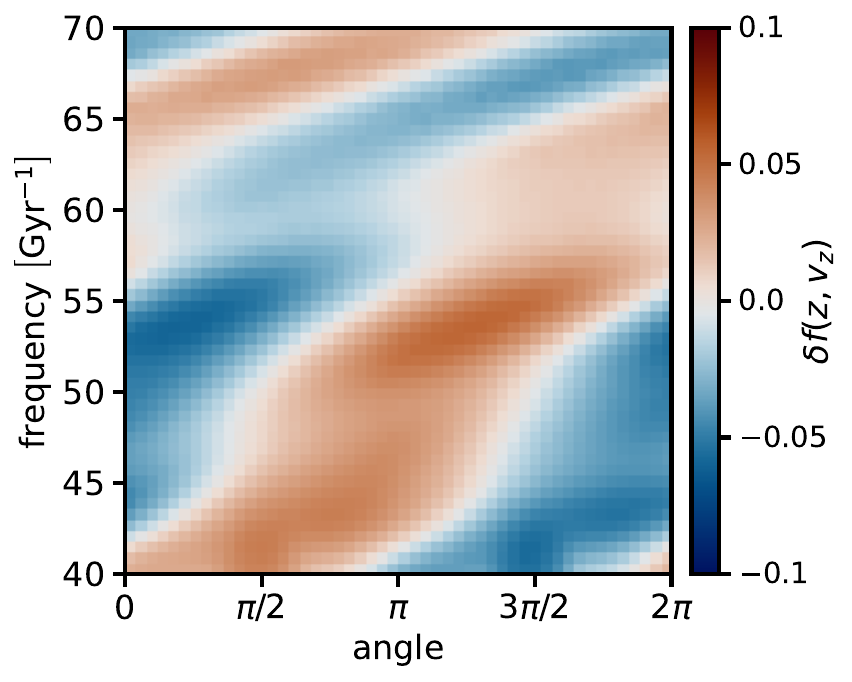}
		\includegraphics[trim=0.25cm 0.5cm 0.5cm
		0.cm,width=0.48\textwidth]{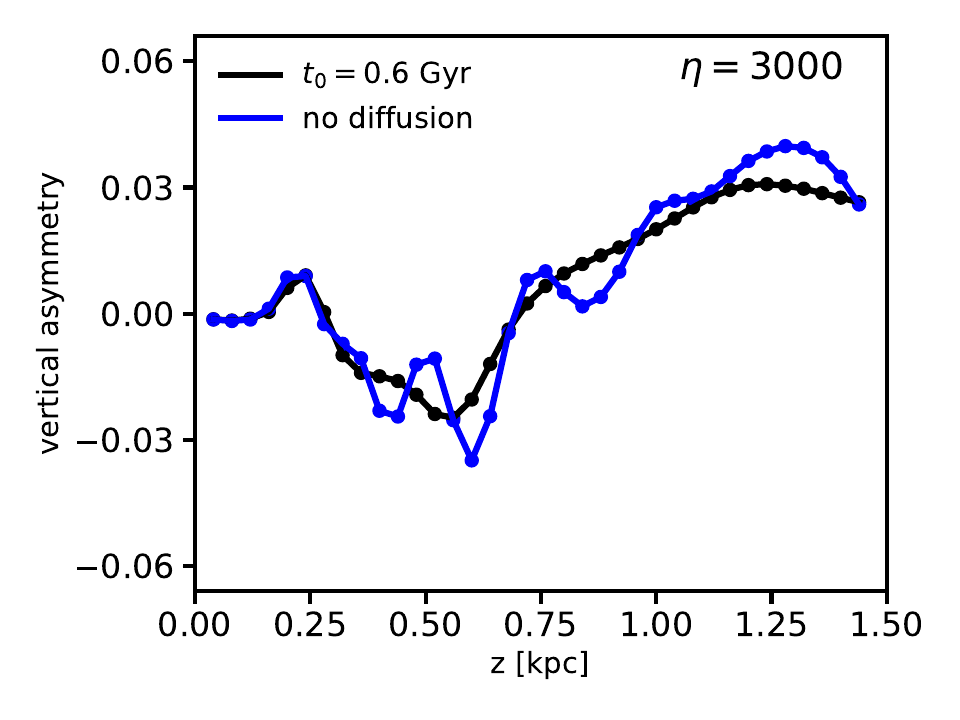}
		\includegraphics[trim=0.25cm 0.5cm 0.5cm
		0.cm,width=0.48\textwidth]{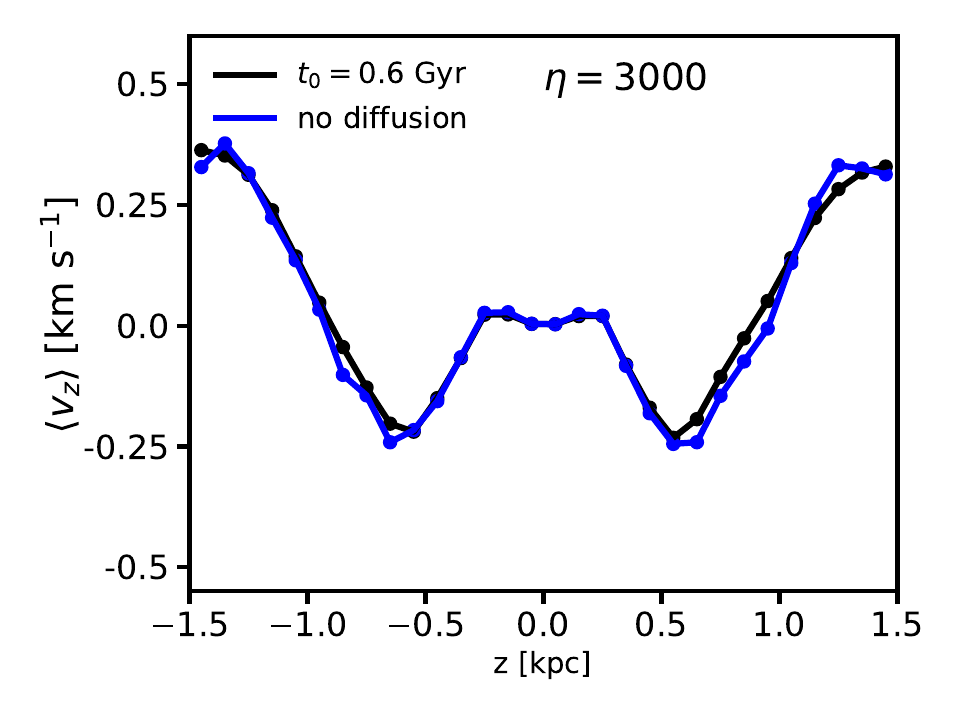}
		\caption{\label{fig:dfexample} An example of a distribution function perturbed by dark subhalos $\left(10^6 < m / M_{\odot}< 10^8\right)$ with an abundance set by $\eta = 3000$. As in Figure \ref{fig:dfexample1992}, the left column does not include diffusion, and the right column includes diffusion.}
	\end{figure*}
	
	Taken together, Figures \ref{fig:forcefig}, \ref{fig:fmaxvshalomass}, and \ref{fig:vimpact} suggest that dark subhalos dominate luminous satellites as the source of external perturbations to the solar neighborhood. This conclusion holds when generating subhalo populations from different random seeds, and for reasonable uncertainties regarding the masses of the luminous satellites and the abundance of subhalos, as determined by $\eta$. However, the vertical forces and $\Delta J_i$'s shown in Figures \ref{fig:forcefig} and \ref{fig:fmaxvshalomass} are not observable. In the next section, we calculate $f\left(z, v_z\right)$, the vertical phase-space distribution of stars in the solar neighborhood subject to perturbations by dark and luminous satellites. 
	
	\subsection{Perturbed distribution functions}
	\label{ssec:impactobs}
	Figure \ref{fig:dfexample1992} shows the perturbation to the distribution function $\delta f\left(z, v_z\right)$ (Equation \ref{eqn:dfpert}) obtained for the same population of dark subhalos whose forces, $\Delta J_i$, and orbital properties are shown in Figures \ref{fig:forcefig}-\ref{fig:vimpact}. For now, we show only the phase-space perturbations caused by subhalos to examine their effects. The top row shows the distribution function in coordinates $\left(z, v_z\right)$. The second row shows the same distribution function in frequency-angle coordinates, $\left(\Omega, \theta\right)$, where $\dot{\theta} = \Omega$, and we calculate the frequencies and angles in the equilibrium potential $\Phi_{\rm{eq}}$. In frequency-angle coordinates, a single instantaneous perturbation manifests as a stripe inclined at an angle $\tau^{-1}$, where $\tau$ is the age of the perturbation. The left column does not include diffusion, while the distribution functions in the right column account for diffusion as discussed in Section \ref{ssec:modeldiffusion}. The bottom row shows two summary statistics of the full phase-space distribution. First, the vertical asymmetry 
	\begin{equation}
		\label{eqn:asym}
		A\left(z\right) = \frac{\rho\left(|z|\right) - \rho\left(-|z|\right)}{\rho\left(|z|\right) + \rho\left(-|z|\right)}
	\end{equation}
	measures deviations in the density $\rho\left(\pm|z|\right)$ above and below the Galactic midplane. Second, we can calculate the first moment of the velocity distribution by integrating over $z$
	\begin{equation}
		\label{eqn:vone}
		v_z^{\left(1\right)}\left(z\right) = \frac{\int v_z f\left(z,v_z\right) d v_z} { \int f\left(z,v_z\right) d v_z}
	\end{equation}
	and compute the mean vertical velocity relative to an observer at $z=0$
	\begin{equation}
		\label{eqn:meanvz}
		\langle v_z\left(z\right) \rangle = v_z^{\left(1\right)}\left(z\right) - v_z^{\left(1\right)}\left(0\right).
	\end{equation}
	In dynamic equilibrium both $A\left(z\right)$ and $\langle v_z\left(z\right) \rangle$ equal zero at all $z$.
	
	In Figure \ref{fig:dfexample}, we show another example of a phase space perturbed by dark subhalos, this time with increased abundance $\left(\eta = 3000\right)$ and initialization of subhalo orbits from a different random seed. We have calculated hundreds of such distribution functions with values of $\eta$ ranging from $500 - 12000$. To illustrate the rich variety of phase spiral morphology that can emerge from stochastic perturbations by dark subhalos, in Appendix \ref{app:B} we provide additional examples. Figures \ref{fig:eta750}, \ref{fig:eta1500}, \ref{fig:eta3000}, \ref{fig:eta6000}, and \ref{fig:eta12000} each show four examples of distribution functions that result from different random seeds and subhalo abundance set by $\eta = 750$, 1500, 3000, 6000, and 12000, respectively. As in Figures \ref{fig:dfexample1992} and \ref{fig:dfexample}, the distribution functions shown in Appendix \ref{app:B} include only perturbations by subhalos. 
	
	Relative to the pattern of oscillations in the  vertical asymmetry and mean vertical velocity caused by the Sagittarius dwarf galaxy \citep[e.g.,][]{BennettBovy21,BennettBovy22}, disturbances caused by subhalos produce more stochastic features in the phase-space distributions, vertical asymmetry, and mean vertical velocity. The perturbed distribution functions in $\left(z, v_z\right)$ coordinates, shown in the first rows of Figures \ref{fig:dfexample1992} and \ref{fig:dfexample} (see also the first and third rows of Figures \ref{fig:eta750} to \ref{fig:eta12000} in Appendix \ref{app:B}) exhibit spiral patterns qualitatively similar to the Gaia snail. In particular, these figures show that subhalos can produce coherent, large-scale correlated features in phase space, properties of distribution functions typically associated with the passage of more-massive satellites. Incorporating the diffusion model described in Section \ref{ssec:modeldiffusion} washes out some of the small-scale structure present in the distribution functions, but our model predicts that coherent spiral structure and fluctuations in the vertical asymmetry and mean vertical velocity should persist until the present day. 
	
	In the second and fourth rows of Figures \ref{fig:dfexample1992} and \ref{fig:dfexample} (see also the second and fourth rows of Figures \ref{fig:eta750} to \ref{fig:eta12000} in Appendix \ref{app:B}), which show the distribution functions in $\left(\Omega, \theta\right)$ coordinates, we observe a complex pattern of stripes with varying inclinations. This is the expected outcome for an ongoing series of perturbations at different times. The stripes in the distribution functions shown in frequency-angle coordinates that include diffusion are systematically more vertical than those that appear without accounting for diffusion, because the perturbations that occurred at times $\tau > t_0$ have since faded away.
	
	\subsection{Statistics of the Gaia snail}
	\label{ssec:dfstats}
\begin{figure}
		\includegraphics[trim=0cm 0cm 0cm
		0cm,width=0.48\textwidth]{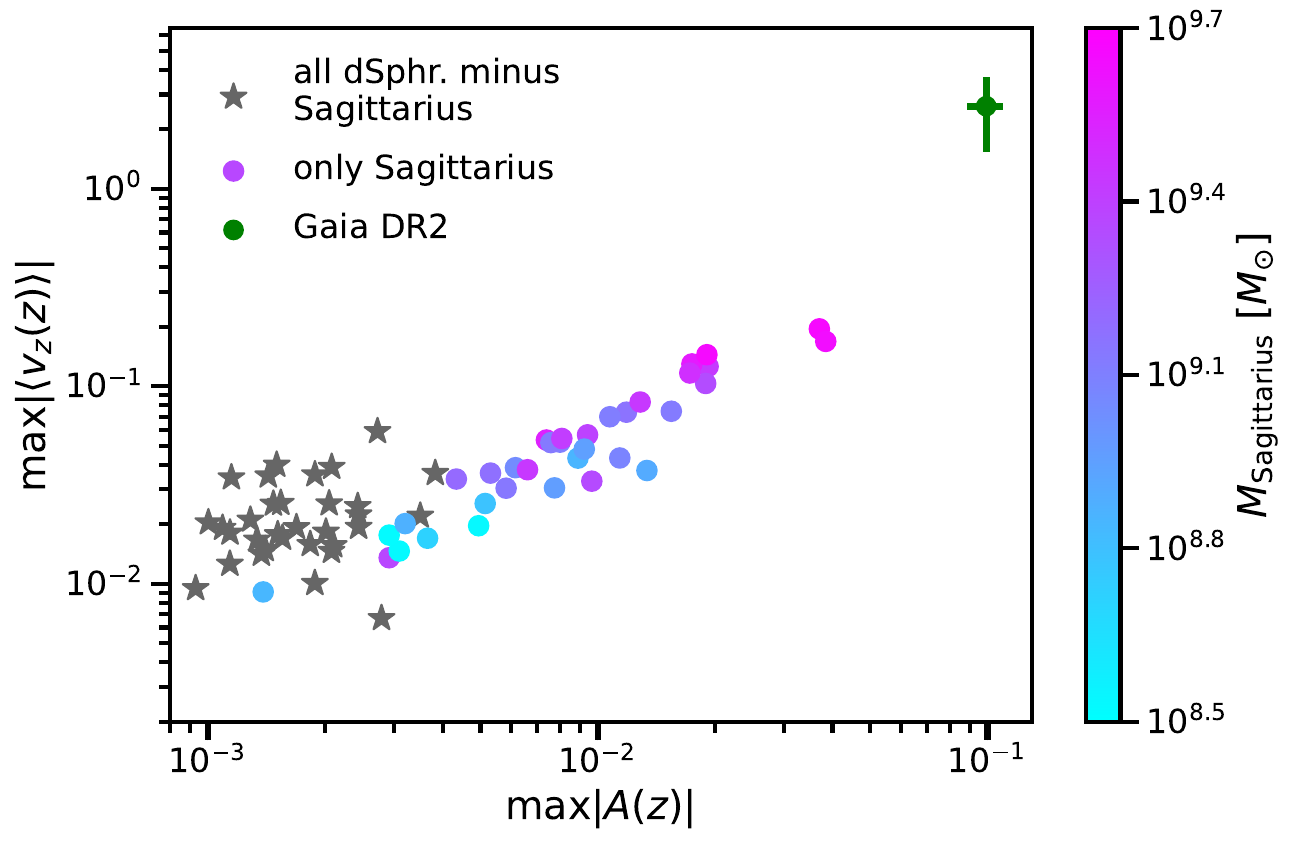}
		\caption{\label{fig:statsdsphr} A comparison between the strength of the perturbations caused by Sagittarius alone, and the rest of the dwarf galaxies listed in Table \ref{tab:luminoussats}. The x-axis shows the maximum amplitude of the vertical asymmetry and the y-axis shows the maximum amplitude of the mean vertical velocity. Each gray star represents the perturbation from 35 realizations of the population of 18 dwarf galaxies listed in Table \ref{tab:luminoussats}, excluding Sagittarius. Colored points show the perturbations caused by Sagittarius alone, with a mass indicated by the color bar. Scatter among the gray stars and colored points comes from uncertainties associated with the halo masses and orbits, as discussed in Section \ref{ssec:dfstats}. The green point shows the maximum amplitude of the mean vertical velocity and vertical asymmetry from Gaia DR2 \citep{BennettBovy21}. }
	\end{figure}

\begin{figure}
		\includegraphics[trim=1cm 0.5cm 0.5cm
		1cm,width=0.48\textwidth]{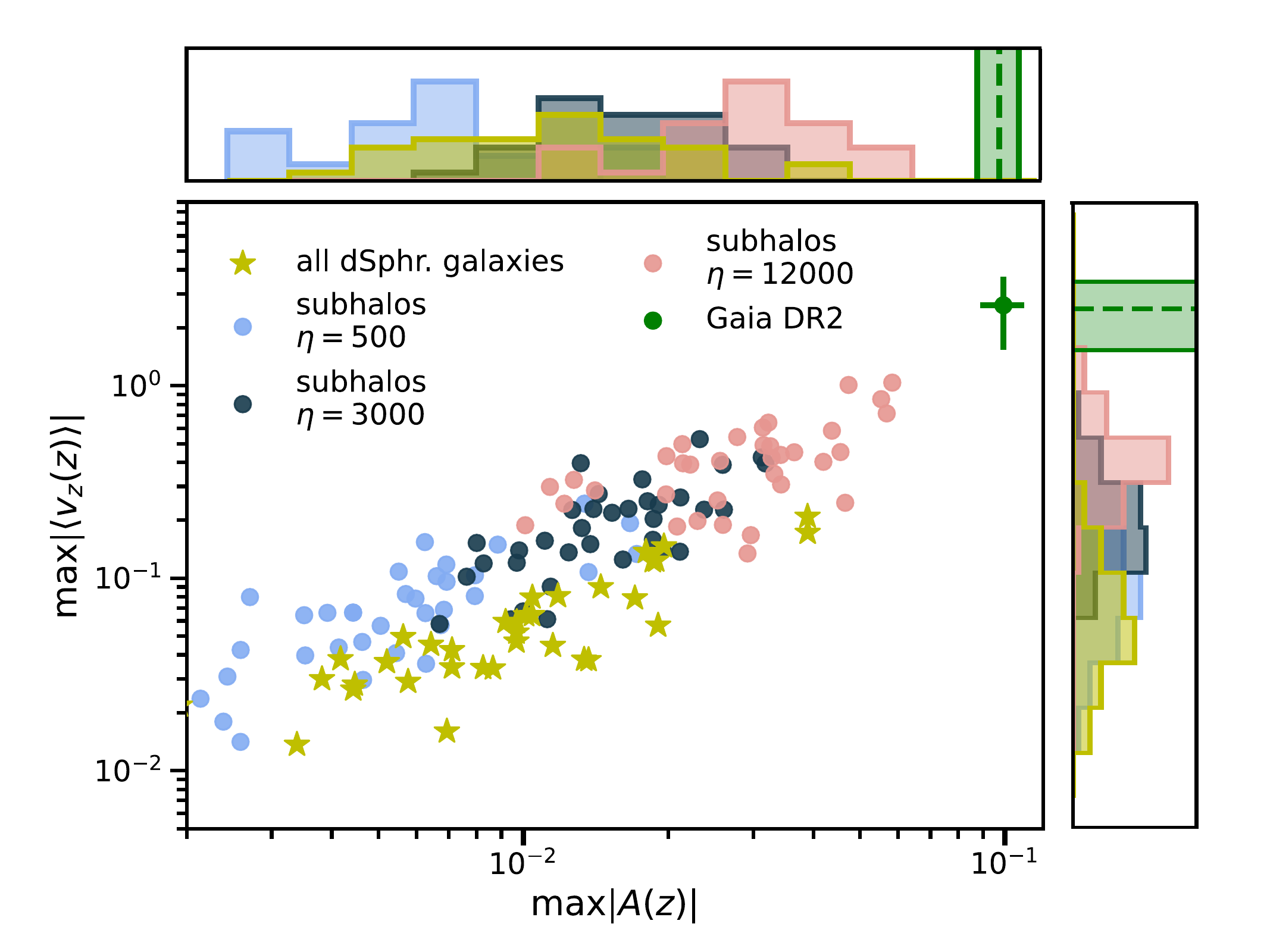}
		\caption{\label{fig:statssplit} The maximum amplitude of the vertical asymmetry (x-axis) and the maximum amplitude of the mean vertical velocity (y-axis) for 35 realizations of subhalo populations (colored points), and 35 realizations of all 18 dwarf galaxies listed in Table \ref{tab:luminoussats}, including Sagittarius (gold stars). The green point shows the maximum amplitude of the mean vertical velocity and vertical asymmetry from Gaia DR2 \citep{BennettBovy19}. Histograms above and below the axes show the marginal distributions of each summary statistic.}
	\end{figure}

\begin{figure}
		\includegraphics[trim=0cm 0cm 0cm
		0cm,width=0.49\textwidth]{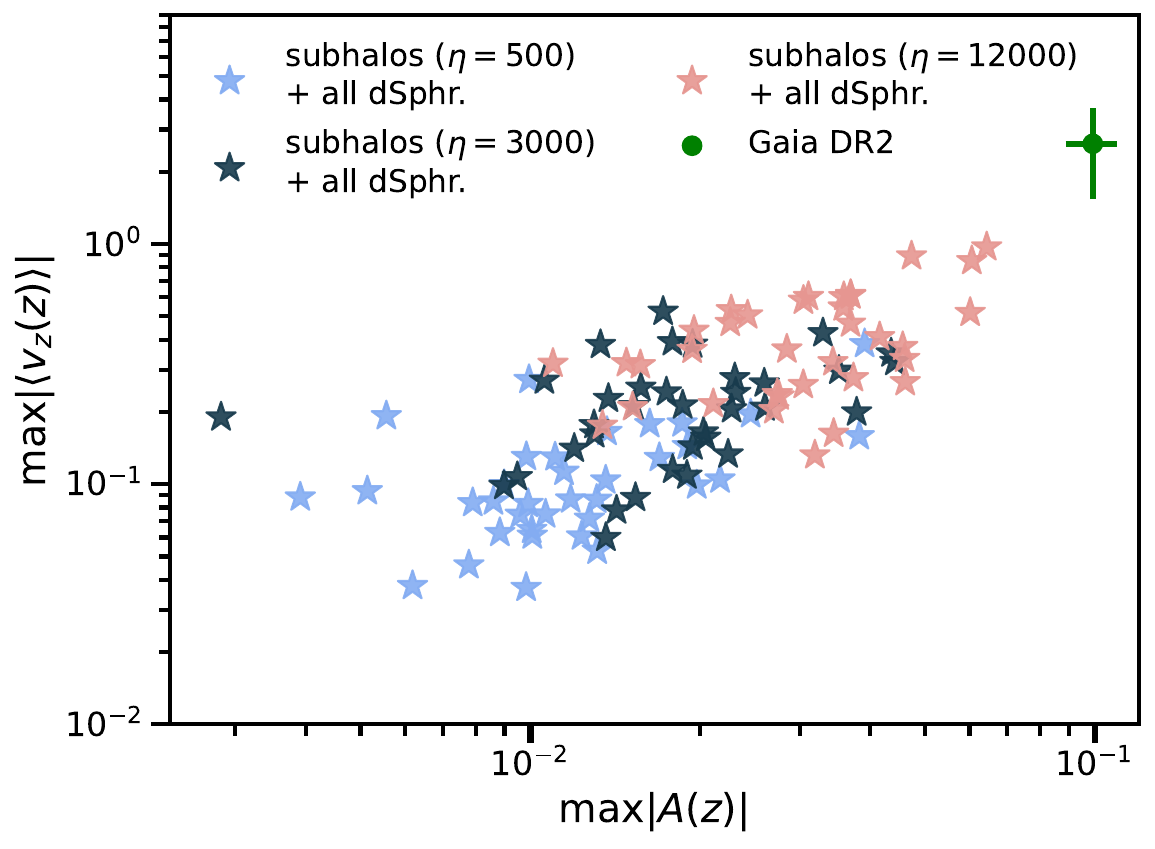}
		\caption{\label{fig:statscombined} The combined effects of subhalos and dwarf galaxies. As in Figure \ref{fig:statsdsphr} and \ref{fig:statssplit}, we show the maximum amplitude of the mean vertical velocity (y-axis) and vertical asymmetry (x-axis) that results from perturbations by a population of perturbers. In this case, we compute the combined effects of dwarf galaxies and subhalo populations with $\eta=500, 3000, 12000$.}
	\end{figure}

	While the properties of the phase spirals that we produce with dark satellites exhibit some qualitatively similar features to the Gaia snail, the strength of the perturbations do not match the data. It is easiest to demonstrate this using summary statistics of the full phase-space distribution: the maximum amplitude of the vertical asymmetry, and the maximum perturbation to the mean vertical velocity. We compare these statistics, as predicted by our model, with the measurements presented by \citet{BennettBovy19}. We make this comparison in three separate parts to illustrate the importance of Sagittarius relative to the rest of the luminous satellites (Figure \ref{fig:statsdsphr}); the importance of subhalos relative to all luminous satellites (Figure \ref{fig:statssplit}); and finally, the combined effect of subhalos and luminous satellites on the dynamics of the solar neighborhood (Figure \ref{fig:statscombined}). All of the results discussed in this section account for diffusion, as outlined in Section \ref{ssec:modeldiffusion}.
	
	Figure \ref{fig:statsdsphr} compares the maximum amplitude of the perturbation caused by Sagittarius (colored points) relative to the other dwarf galaxies listed in Table \ref{tab:luminoussats}, represented as gray stars. Each star shows the perturbation caused by all satellites listed in the table, minus Sagittarius. The scatter among the points and stars derives from uncertainties associated with the satellite orbits and masses, which we calculate as follows: for each simulated population of dwarf galaxies, we sample uncertainties in their present-day positions and proper motions using the kinematic data from the Local Volume Database \citep{Pace++24}. We integrate these orbits backwards in the {\tt{MWPotential2014}} in {\tt{galpy}} while also rescaling the mass of the Milky Way's host dark-matter halo by random factors $0.7 - 1.3$ to account for uncertainties in the orbits connected to the gravitational potential of the galaxy, which is dominated by the dark-matter halo. To account for the effects of tidal stripping, we rescale the mass of each dwarf galaxy by a random factor between $0.1 - 1.0$. 
	
	We see that Sagittarius causes stronger perturbations to the dynamics of the solar neighborhood than the combined effect of all other dwarf galaxies listed in Table \ref{tab:luminoussats}. This is consistent with previous work identifying Sagittarius as the most relevant known perturber in the context of the Gaia snail \citep[e.g.,][]{Banik++22}. We calculate a weaker response by the disk to the passage of Sagittarius than reported by \citet{BennettBovy21} due to the assumed model of the stellar distribution function. \citet{BennettBovy21} used an isothermal distribution function, for which the velocity dispersion, $\sigma_v$, is approximately independent of height, while for Equation \ref{eqn:liwidrowdf}, $\sigma_v$ increases approximately linearly with $z$.  
	
	We now compare the strength of perturbations caused by all luminous satellites as a single population with those produced by dark subhalos. In Figure \ref{fig:statssplit}, we show the maximum amplitude of the vertical asymmetry and mean vertical velocity produced by subhalos (colored points) relative to dwarf galaxies (gold stars). We account for uncertainties in the dwarf galaxy orbits and masses in the same way as described for Figure \ref{fig:statsdsphr}. One should interpret the range of $\eta$ explored in the figure as bracketing the range of uncertainties associated with both the overall number of subhalos, and which mass definition (bound mass versus infall mass) most appropriately captures their dynamic effects. Populations of dark subhalos with $\eta \gtrsim 3000$ impart stronger perturbations to both the mean vertical velocity and vertical asymmetry than the luminous satellites. For $\eta = 500$, corresponding to the low end of the expected subhalo abundance and a mass definition that minimizes their dynamic effects (the gravitationally bound mass today), subhalos and luminous satellites impart perturbations of a similar strength. Thus, for most reasonable priors on the abundance of subhalos and the mass definition assigned to them, we find subhalos impart stronger perturbations to the dynamics of the Gaia snail than Milky Way's known population of luminous satellites, including Sagittarius. 
	
	Considering models with only dark subhalos, we find that both statistics scale approximately as $\sqrt{\eta}$; specifically, $\max |v_z\left(z\right)| = A \sqrt {\eta/1000}$ and $\max |A\left(z\right)| = B \sqrt {\eta/1000}$ with $A = 0.10 \pm 0.03 \ \rm{km} \ \rm{s^{-1}}$ and $B = 0.008 \pm 0.002 $. Including dwarf galaxies causes additional scatter around these relations, depending mainly on the assumptions for the orbit and mass of Sagittarius. 
	
	We now consider the effects of subhalos and dwarf galaxies as a single population. Figure \ref{fig:statscombined} shows the perturbations to the phase space from the combined effects of luminous satellites and subhalos, with the subhalo abundance increasing from $\eta = 500$ to $12,000$. None of configurations of combined luminous and dark satellites can simultaneously match the strength of the perturbation to the mean vertical velocity and the vertical asymmetry unless we have $\eta > 12,000$, roughly 5-10 times greater than the number of perturbing satellites we estimate from {\tt{galacticus}}. 
	
	\section{Discussion and conclusions}
	\label{sec:conclusions}
	We have carried out numerical simulations to examine the effect of perturbations by a population of low-mass ($10^6 M_{\odot} < m < 10^8 M_{\odot}$) Galactic subhalos on the dynamics of the Gaia phase spiral. Our model for the dark subhalo population is based on evolving halo merger trees in a Milky Way-like potential using a semianalytic model of structure formation. Using an approximate analytic framework for calculating the response of stars near the Galactic midplane to disturbances caused by low-mass dark satellites, we calculate the perturbations to the stellar distribution function caused by these dark halos, and examine how the signatures from past disturbances caused by dark subhalos affect various summary statistics, such as the mean vertical velocity and vertical number count asymmetry. Our modeling framework also includes a phenomenological treatment of diffusion by gravitational scattering between stars and giant molecular clouds, which damps perturbations to the distribution function from disturbances older than $\sim 0.6$ Gyr. We summarize our main results as follows: 
	\begin{itemize}
		\item Dark subhalos alone do not produce strong enough perturbations to the dynamics of the solar neighborhood to explain all properties of the Gaia snail. However, they constitute a more probable source of significant perturbation $\langle |\Delta J| \rangle / \langle J_{\rm{eq}}\rangle \gtrsim 10^{-2}$ to the dynamics of the solar neighborhood than the less-numerous and more-distant dwarf galaxies. 
		\item Assuming a subhalo abundance predicted by CDM, we expect fluctuations in the vertical number count asymmetry and the mean vertical velocity on the order of $1-3 \%$ and $0.1 - 0.4 \ \rm{km} \ \rm{s^{-1}}$, respectively. 
		\item Ongoing  perturbation by dark subhalos produces stochastic features in the vertical $\left(z, v_z\right)$ distribution function. In frequency-angle coordinates, these features manifest as numerous horizontal stripes with varying inclinations. 
	\end{itemize}
	
	Our results extend early work by \citet{Feldmann++15} and \citet{Buschmann++18}, who motivated the study of the Milky Way's subhalos using precise kinematic measurements from Gaia. The topic of perturbations by subhalos was also recently addressed by \citet{GarciaConde++24} using N-body simulations, although we note that the $10^9 M_{\odot}$ perturbers they identified would likely host luminous dwarf galaxies based on the current understanding of the galaxy-halo connection \citep{Nadler++20}. 
	
	We have built on these analyses by estimating the response of the Galactic disk to external perturbations using a computationally tractable one-dimensional model for the local phase space. We study the properties of phase-space spirals at much higher resolution than one can attain in N-body simulations, while exploring a range of possible subhalo orbital configurations, abundances, and masses, while resolving features of the distribution function that reveal interactions with multiple dark halos with masses down to $10^6 M_{\odot}$. Subhalos impart a stochastic pattern of oscillations in the vertical asymmetry and mean vertical velocity with varying wavelength and amplitude. The appearance of the phase-space distribution in frequency-angle coordinates, subject to perturbations by subhalos, exhibits a complex morphology, with numerous stripes arrayed at various inclinations. This feature of the distribution function, if observed, could be interpreted as a signature of ongoing perturbation at many different times. Previous studies have identified some hints of multiple perturbation events in the Gaia snail \citep[e.g.][]{Frankel++23,Antoja++23}. 
	
	When considering the effects of satellite galaxies, many works have identified the most-massive perturbers as the dominant source of dynamic perturbation. Indeed, the strength of a perturbation increases with the mass of the perturber, on average, and therefore the average strength of a perturbation by a dwarf galaxy exceeds the average strength of the pertubation caused by a dark subhalo. However, we know (approximately) the orbits of the Milky Way's luminous satellites, whereas we do not know the orbits of the more ubiquitous dark subhalos. When we allow for the possibility that their orbits bring them in close proximity to the solar neighborhood, we find that subhalos constitute a more probable source of strong perturbation than the less-abundant and more-distant luminous satellites. In particular, our analysis suggests that dark subhalos contribute to the disequilibrium measured by Gaia to at least the same level as Sagittarius, and reliably cause stronger perturbations than Sagittarius, and all other dwarf spheroidal galaxies around the Milky Way, for $\eta \gtrsim 3000$. This value of $\eta$ corresponds to 924 halos with bound masses in the range $10^7 - 10^{8} M_{\odot}$ passing within 50 kpc of the Galactic center in the past 2.4 Gyr.
	
	Despite the possibility that subhalos can affect the stellar dynamics that manifests in the Gaia snail, the abundance necessary to match the strength of the perturbations measured by Gaia exceeds the number of dark satellites predicted by CDM by a factor $5-10$, based on our simulations with ${\tt{galacticus}}$. The implied abundance of subhalos, assuming satellites alone triggered the formation of the snail, is also in tension with recent inferences of the Milky Way's subhalo mass function from stellar streams \citep{BanikBovy++21} and the abundance of dwarf galaxies \citep{Nadler++20,Dekker++22}. These results suggest another mechanism, likely acting in combination with dark and luminous satellites, gave rise to the Gaia snail. 
	
	In this work, we have used a one-dimensional model developed by \citet{BennettBovy21} to examine the kinematic signatures of dark subhalos in the $z - v_z$ phase-space distribution. As shown in Appendix \ref{app:A}, this approximation works quite well, even for multiple perturbers. However, phase spirals appear in higher dimensions, and become especially prominent when considering motion along the radial and azimuthal directions \citep{Antoja++23}, in different regions of the galaxy \citep{Hunt++21}, or when selecting on stellar angular momentum, age, or chemical composition \citep{Gandhi++22,Frankel++23,Frankel++24}. Extending the one-dimensional model used in this work to predict the morphology of phase spirals in higher dimensions --- radial phase mixing \citep[e.g.][]{Hunt++24}, for example --- could aid in disentangling multiple sources of dynamic perturbation.
	
	This analysis has focused on the solar neighborhood and the Gaia snail, but our results suggest disturbances by dark-matter substructure should manifest in phase spirals throughout the Galaxy \citep[e.g.][]{Hunt++21,Frankel++23}. Unfortunately, we expect the kinematic signatures associated with subhalos to have short lifetimes. As noted by \citet{Tremaine++23}, gravitational scattering against giant molecular clouds, once it begins, rapidly diffuses the phase mixing signature of a past disturbance. We have included a model for diffusion in this work, and show that some perturbative signatures associated with dark subhalos should persist to the present day. Detecting these signatures would provide independent evidence for the existence of dark subhalos in our Galaxy---one of the fundamental predictions of CDM, and other theories with particle dark matter. 
	
	\section*{Acknowledgments}
	We thank Alex Drlica-Wagner and Gerry Gilmore for encouraging discussions. We also thank Ethan Nadler for sharing scripts to implement the galaxy-halo connection model presented in \citet{Nadler++20}. 
	
	DG acknowledges support for this work provided by the Brinson Foundation through a Brinson Prize Fellowship grant, and by a HQP grant from the McDonald
	Institute (reference number HQP 2019-4-2). JB acknowledges financial support from the Natural Sciences and Engineering Research Council of Canada (NSERC; funding reference number RGPIN-2020-04712).
	NF acknowledges the support of the Natural Sciences and Engineering Research Council of Canada (NSERC), funding reference numbers 568580 and RGPIN-2020-03885, and partial support from an Arts \& Sciences Postdoctoral Fellowship at the University of Toronto. This work used computational and storage services associated with the Hoffman2 Cluster which is operated by the UCLA Office of Advanced Research Computing’s Research Technology Group. 
	
	\section*{Software} This work made use of {\tt{astropy}}:\footnote{http://www.astropy.org} a community-developed core Python package and an ecosystem of tools and resources for astronomy \citep{astropy:2013, astropy:2018, astropy:2022}; {\tt{numpy}} \citep{numpy}; {\tt{galacticus}}\footnote{https://github.com/galacticusorg/galacticus/wiki} \citep{Benson12}; and {\tt{galpy}} \citep{Bovy15}. The code developed for this project, {\tt{darkspirals}}, is available on github\footnote{https://github.com/dangilman/darkspirals}.

	
	\bibliography{bibliography}
	\bibliographystyle{aasjournal}
	
	\appendix
	\section{Quantifying the accuracy of the approximation for $\Delta J$}
	\label{app:A}
	\begin{figure*}
		\includegraphics[trim=0cm 0.5cm 0cm
		0cm,width=0.48\textwidth]{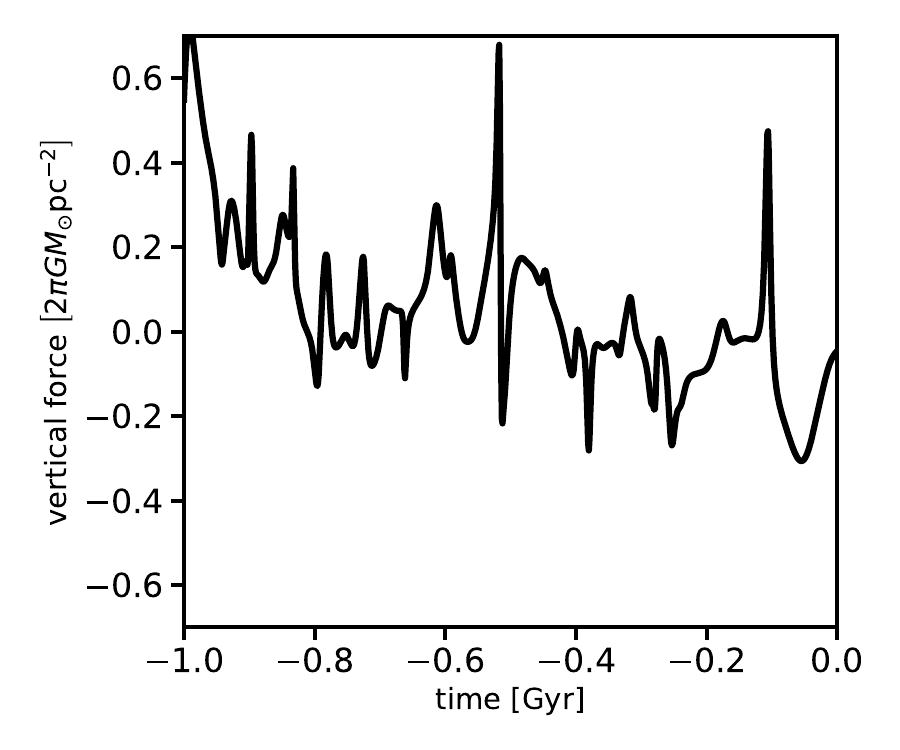}
		\includegraphics[trim=0.5cm 0.5cm 0.5cm
		2cm,width=0.48\textwidth]{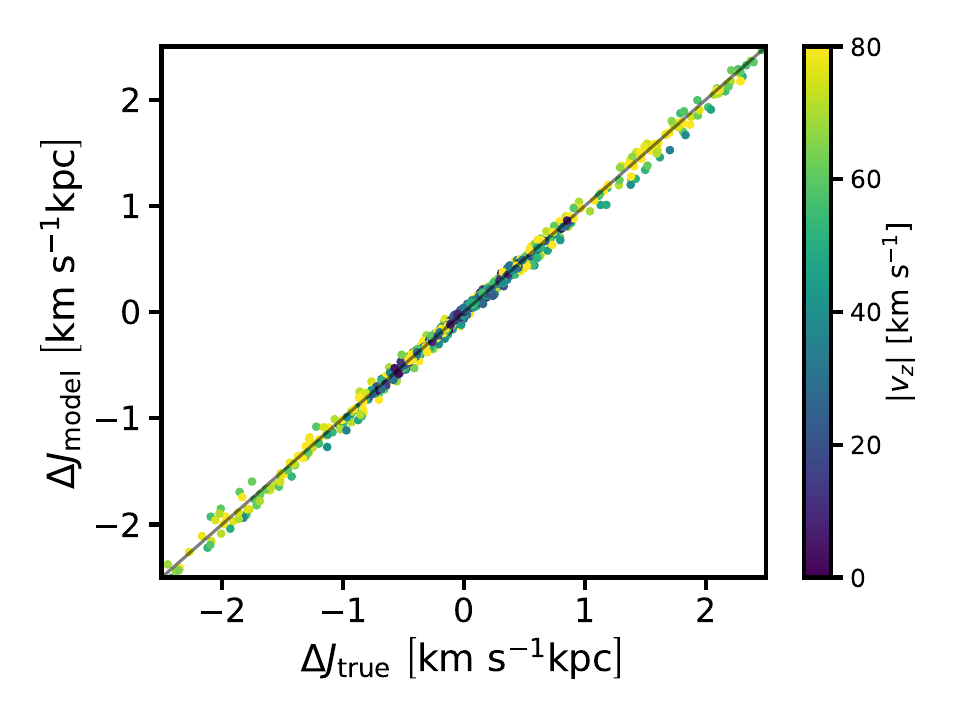}
		\caption{\label{fig:djtest} Vertical perturbing forces as a function of time (left) and a comparison between the exact and model-predicted changes to the vertical action (right). Points in the right panel represent samples drawn uniformly from the phase-space area color coded by their vertical velocity $v_z$.}
	\end{figure*}
	\begin{figure*}
		\includegraphics[trim=0.5cm 0.5cm 0.5cm
		1cm,width=0.33\textwidth]{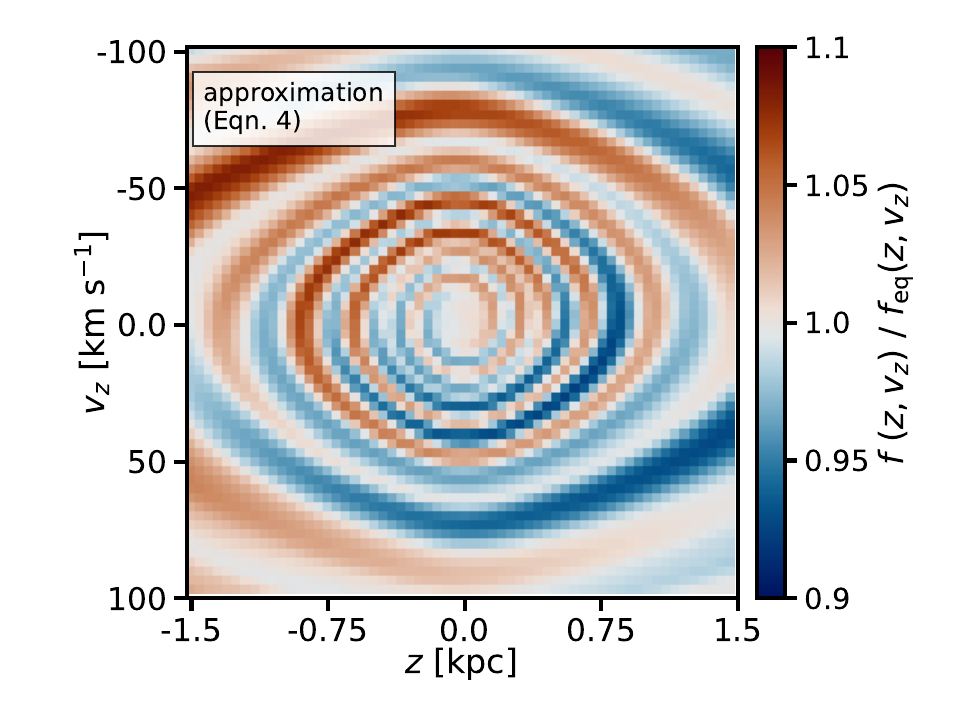}
		\includegraphics[trim=0.5cm 0.5cm 0.5cm
		1cm,width=0.33\textwidth]{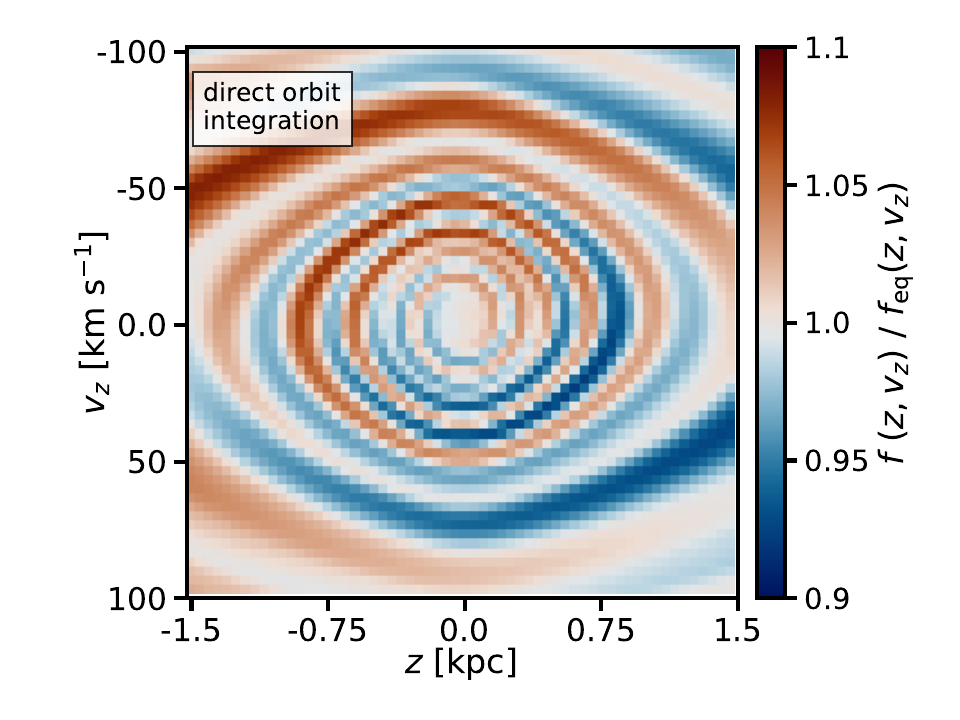}
		\includegraphics[trim=0.5cm 0.5cm 0.5cm
		1cm,width=0.33\textwidth]{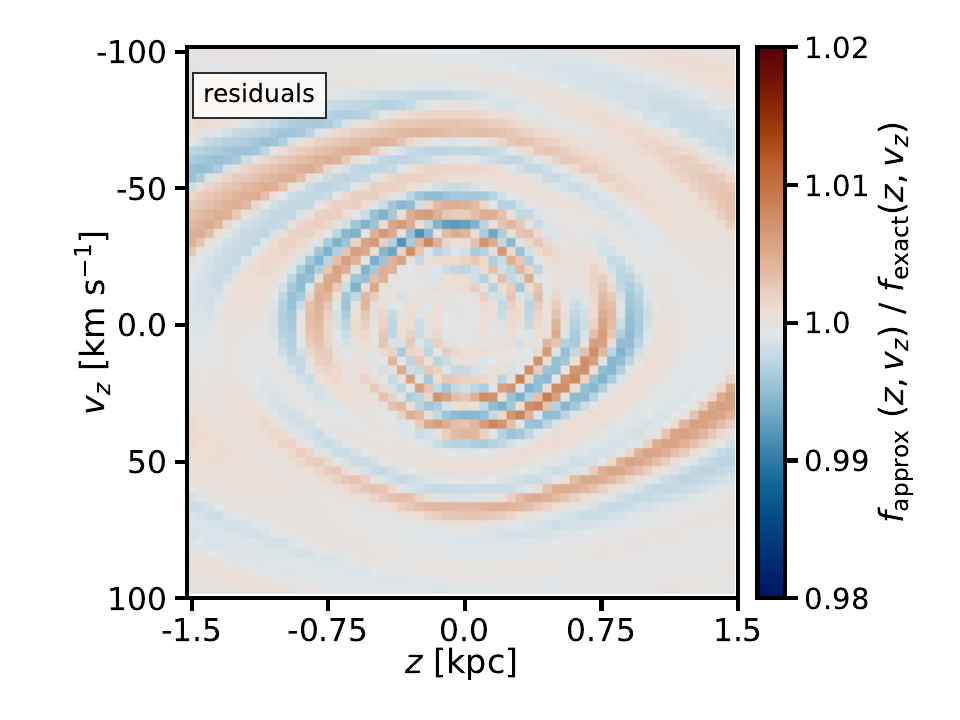}
		\caption{\label{fig:dftest} The distribution function produced from the series of perturbing forces shown in Figure \ref{fig:djtest} using the approximation in Equation \eqref{eqn:dji} (left), direct orbit integration to compute $\Delta J$ (center), and residuals between the two (right). Typical systematic errors in the distribution function incurred from the simplifying assumptions associated with the approximation in Equation \ref{eqn:dji} are at the level of $1-2\%$.}
	\end{figure*}
	Equation \eqref{eqn:dji} enables a significantly more efficient calculation of perturbed distribution functions than direct orbit integration but involves certain simplifying assumptions. In particular, Equation \eqref{eqn:dji} assumes that satellites impart small perturbations to stellar orbits, such that we can calculate the change in the vertical action by integrating along an unperturbed trajectory. Equation $\eqref{eqn:dji}$ also breaks down for $v_z = 0$, corresponding to stars sitting in the Galactic midplane with zero vertical motion (although this possibility is unlikely to be realized in nature). 
	
	To assess the accuracy of Equation \eqref{eqn:dji}, we compare the prediction for $\Delta J$ given by the model with exact calculation of $\Delta J$ by direct orbit integration in a live potential. The perturber population includes two objects with the same orbits as Sagittarius and Tucana III, to which we assign masses $10^{9} M_{\odot}$ and $10^{8} M_{\odot}$, respectively, and a population of dark subhalos $\left(10^7 - 10^8 M_{\odot}\right)$ with an abundance set by $\eta = 2500$. For the orbits in the live potential, we calculate $\Delta J$ according to Equation \eqref{eqn:dji}, and again by integrating the orbit of test particles in the live potential with the full population of satellites. After computing $\Delta J$ with both methods, we evaluate the distribution function in Equation \eqref{eqn:liwidrowdf} at the perturbed coordinates. 
	
	The left panel of Figure \ref{fig:djtest} shows the vertical force sourced by the satellite population used for this test. The right panel of Figure \ref{fig:djtest} compares the approximation for $\Delta J$ given by Equation \eqref{eqn:dji} with the exact calculation of $\Delta J$ obtained by direct orbit integration. We find that $65 \%$ of samples have an absolute error on $\Delta J$ less than $10 \%$, and $82 \%$ of samples have an absolute error less than $20\%$. To see how these systematic uncertainties propagate to the distribution function, Figure \ref{fig:dftest} shows the perturbation to the distribution function obtained using the methods in this paper (left), the exact calculation with direct orbit integration (center), and the residuals (right). Structure in the residuals appears due to the simplifying assumptions in the methods we use to calculate the change in the vertical action. However, these differences are at the level of a few percent in the distribution function, which is accurate enough to ensure this source of systematic uncertainty does not affect our main conclusions.  
	
	\section{Additional visualization of perturbed distribution functions}
	\label{app:B}
\begin{figure*}
		\includegraphics[trim=1.5cm 7.5cm 3.cm
		0cm,width=0.95\textwidth]{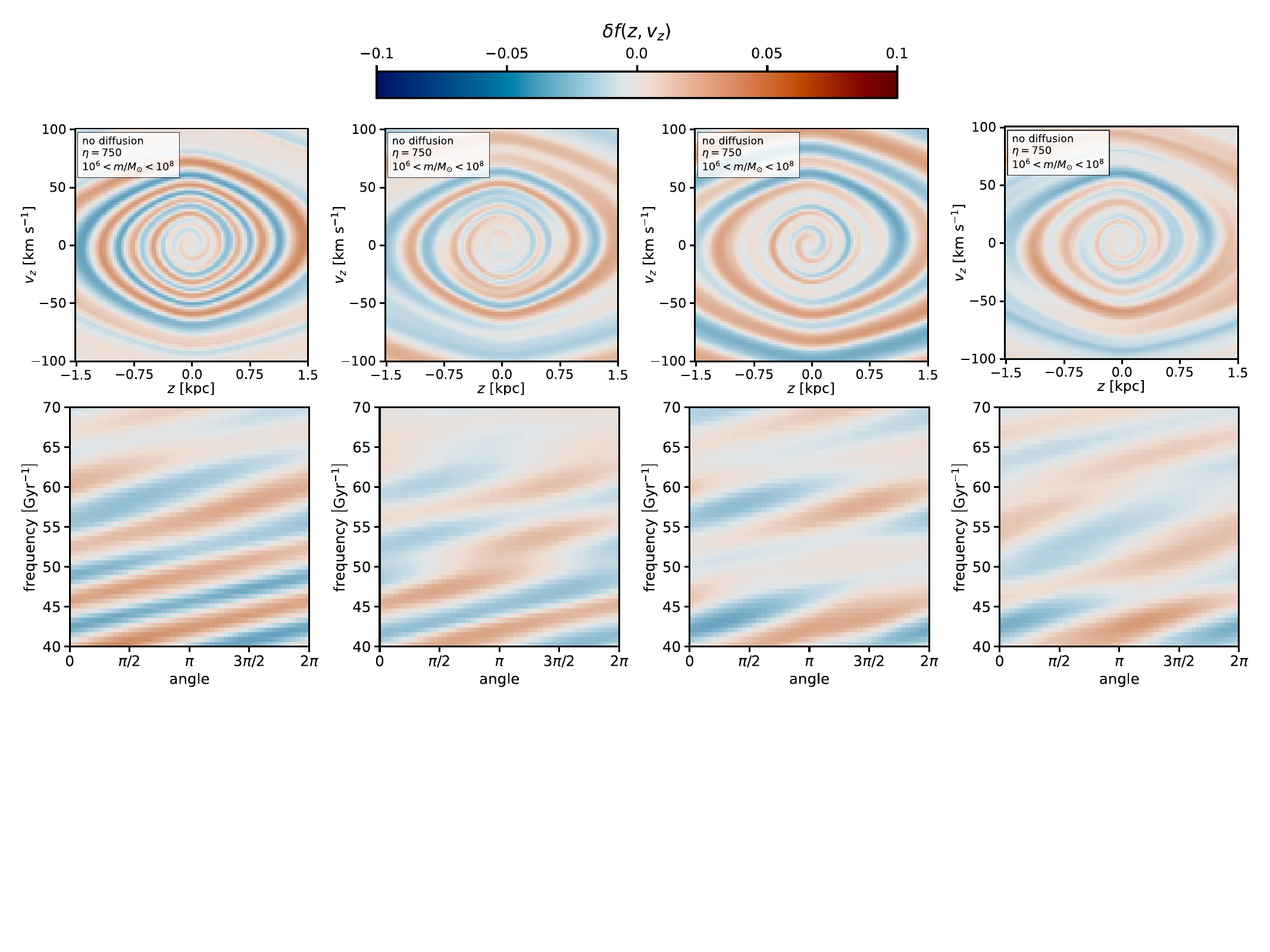}
		\includegraphics[trim=1.5cm 10.cm 3.cm
		0cm,width=0.96\textwidth]{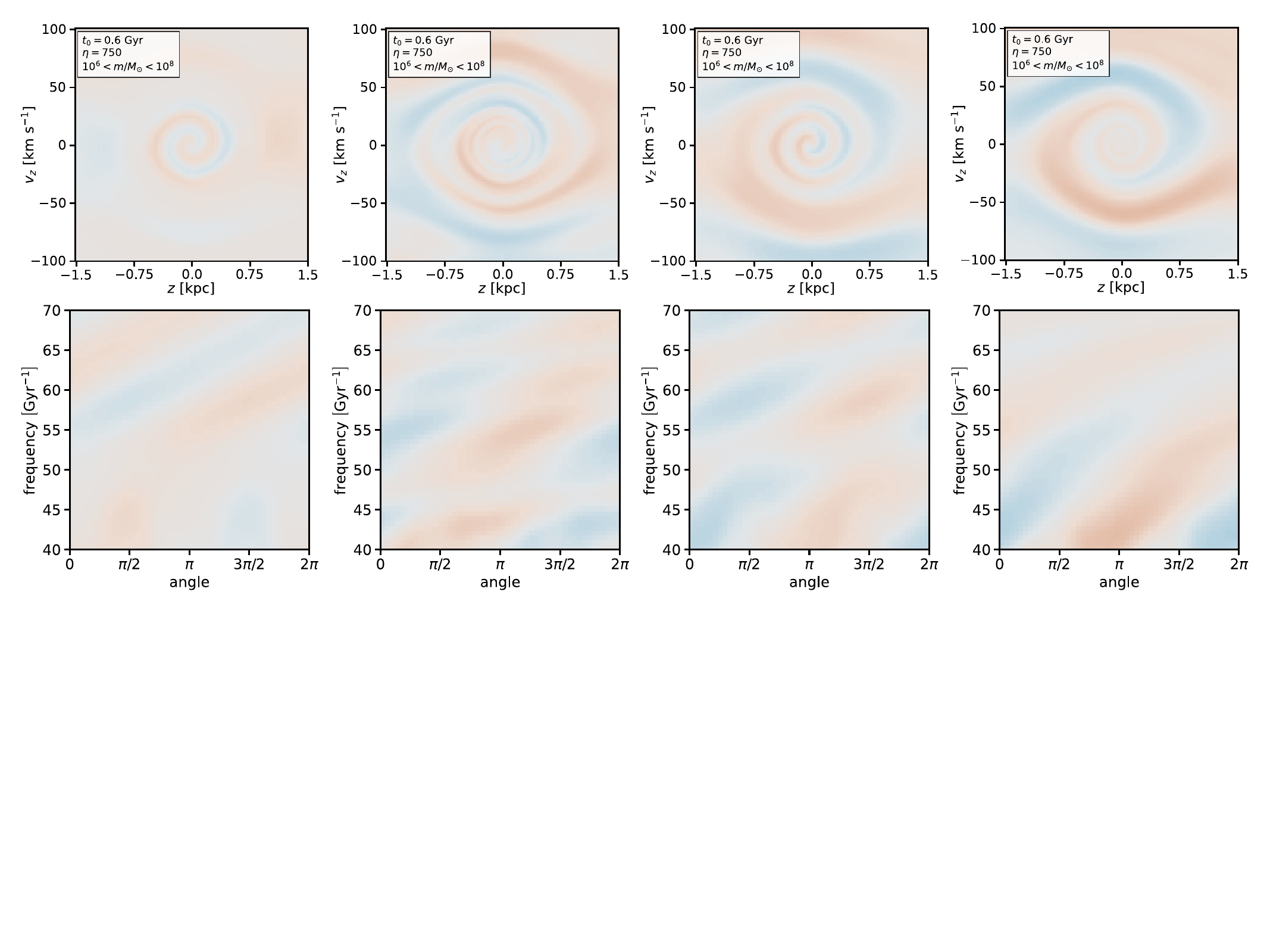}
		\caption{\label{fig:eta750} Perturbed distribution functions produced from four realizations of dark subhalos from different random seeds. Only perturbations by dark subhalos are included in the model, and the abundance is set by $\eta=750$. The first and third rows show the perturbation to the distribution function in $\left(z,v_z\right)$ coordinates, and the second and fourth rows show the perturbation in $\left(\Omega, \theta\right)$ coordinates (frequency and angle, respectively). The top two rows do not include diffusion, and the bottom two rows show the same perturbations with diffusion included.}
	\end{figure*}

\begin{figure*}
		\includegraphics[trim=1.5cm 7.5cm 3.cm
		0cm,width=0.95\textwidth]{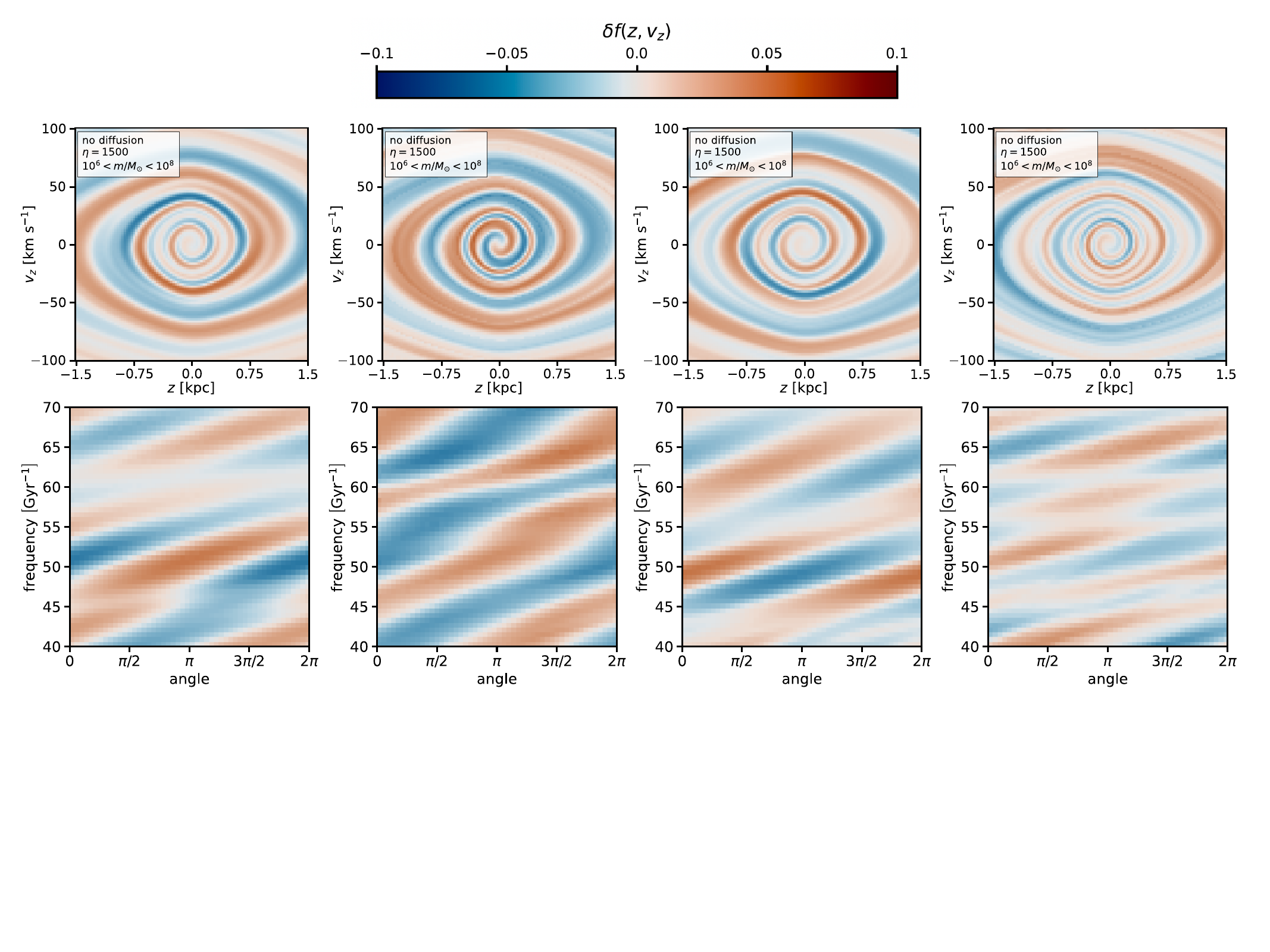}
		\includegraphics[trim=1.5cm 10.cm 3.cm
		0cm,width=0.96\textwidth]{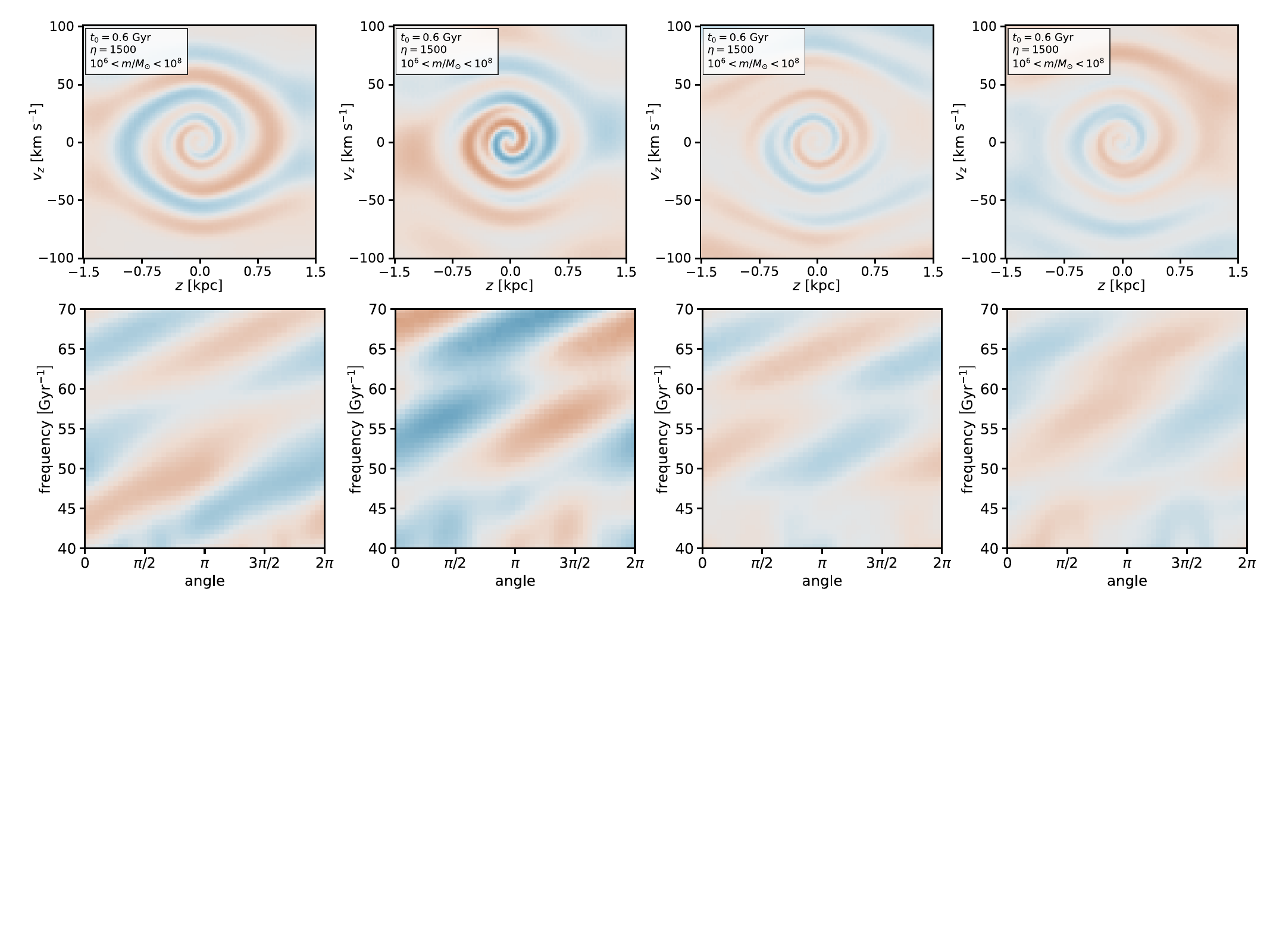}
		\caption{\label{fig:eta1500} The same as Figure \ref{fig:eta750}, but with subhalo abundance set by $\eta=1500$.}
	\end{figure*}

\begin{figure*}
		\includegraphics[trim=1.5cm 7.5cm 3.cm
		0cm,width=0.95\textwidth]{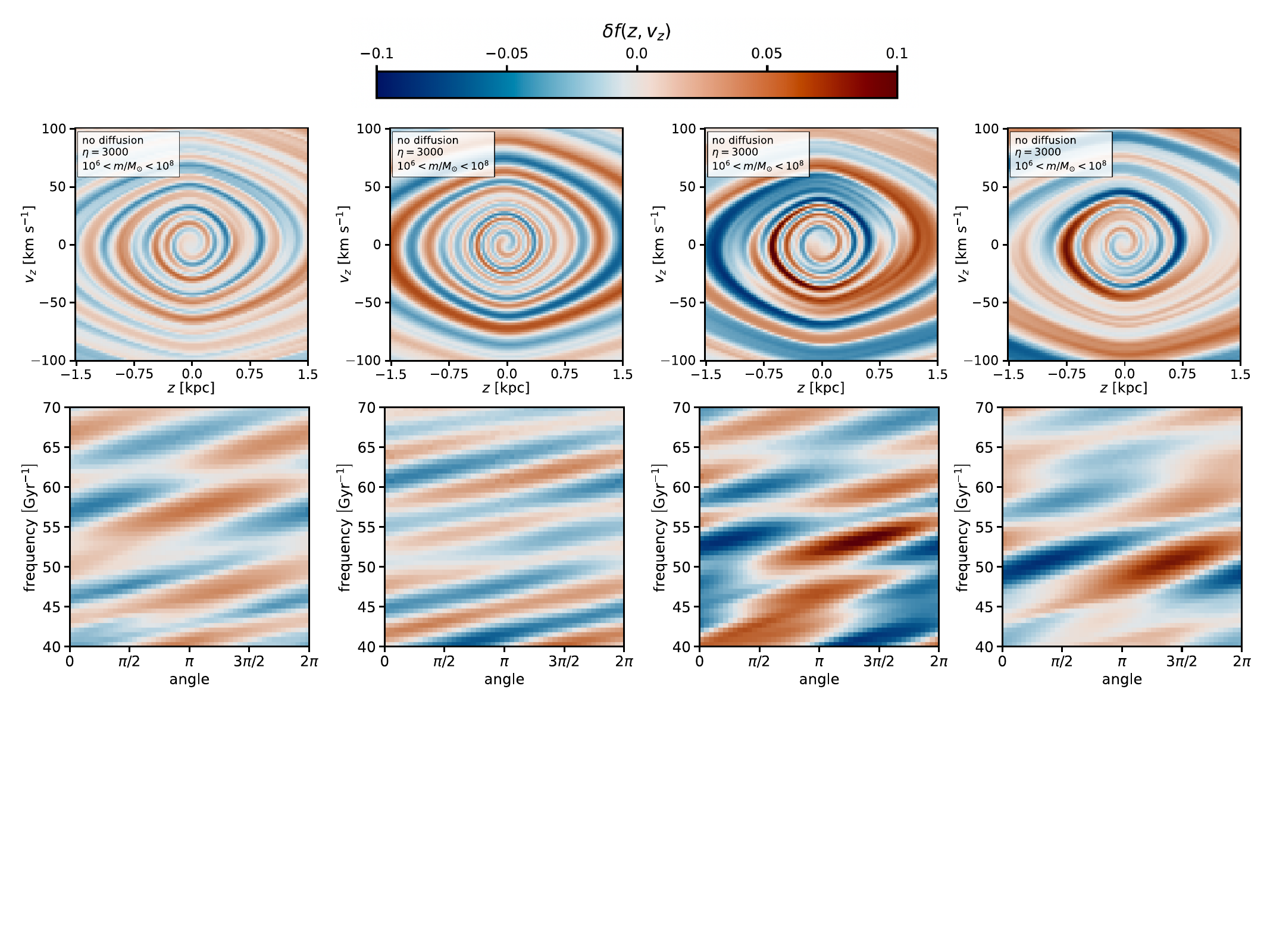}
		\includegraphics[trim=1.5cm 10.cm 3.cm
		0cm,width=0.96\textwidth]{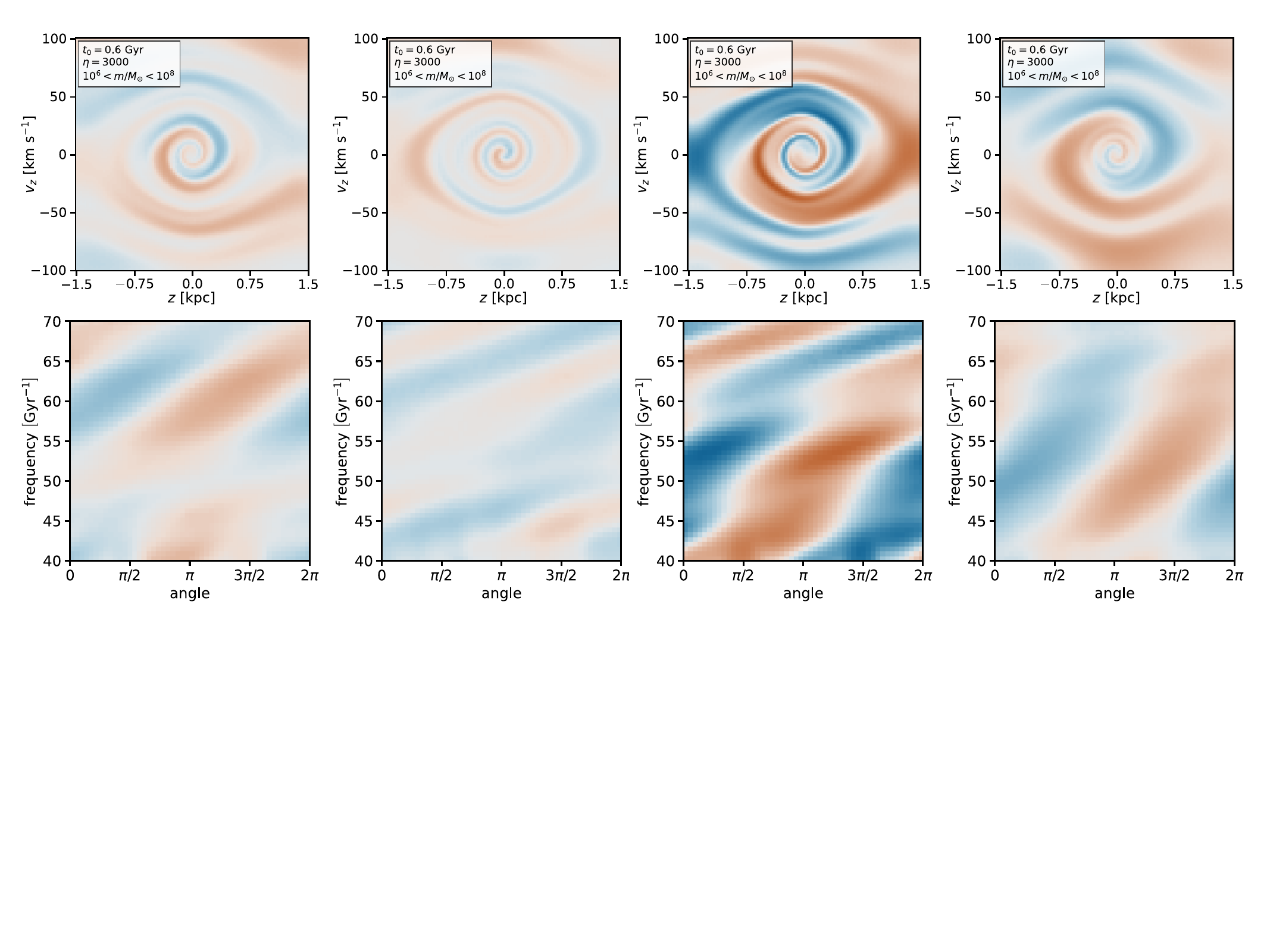}
		\caption{\label{fig:eta3000} The same as Figures \ref{fig:eta750} and \ref{fig:eta1500}, but with subhalo abundance set by $\eta=3000$.}
	\end{figure*}

\begin{figure*}
		\includegraphics[trim=1.5cm 7.5cm 3.cm
		0cm,width=0.95\textwidth]{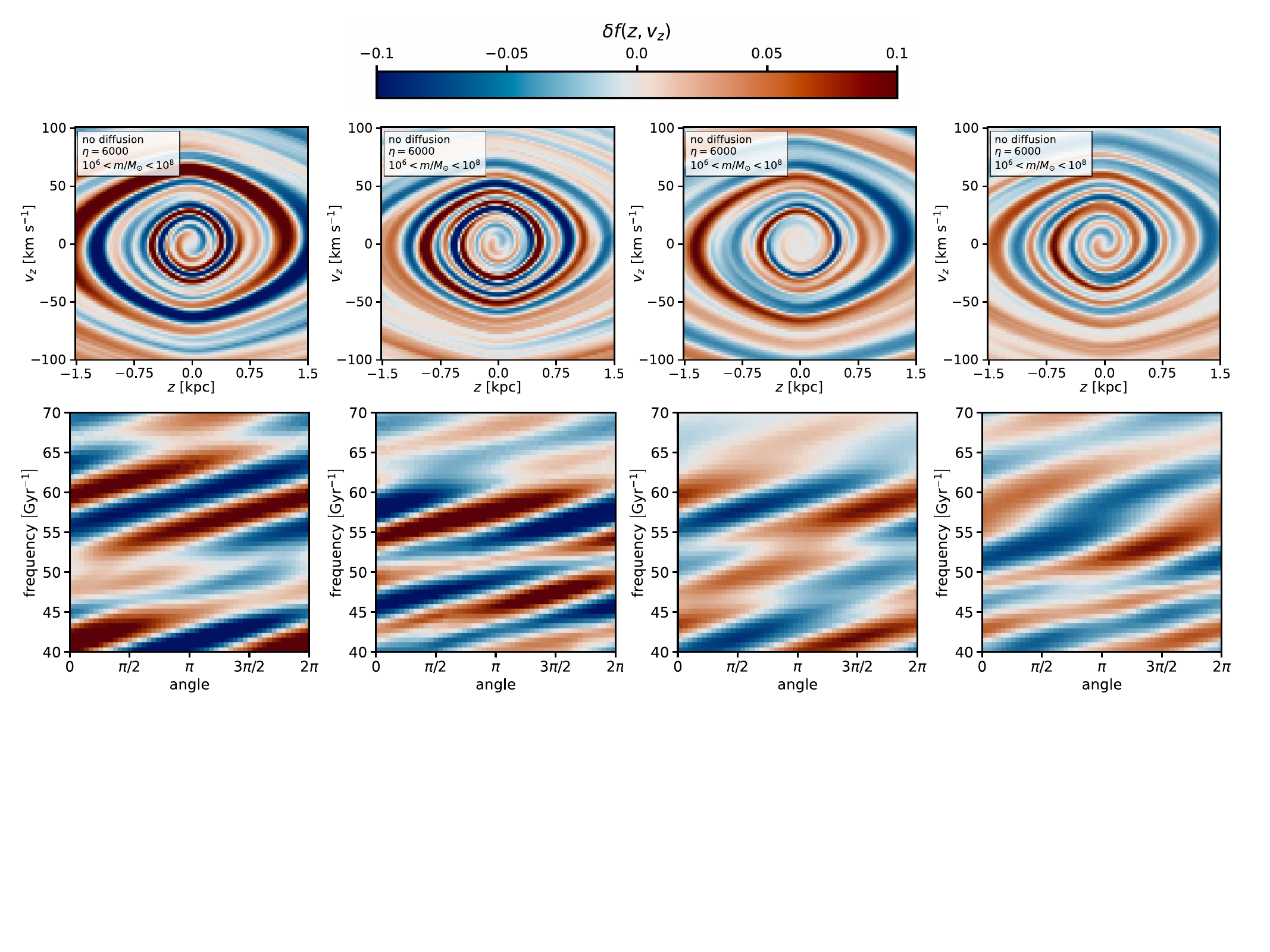}
		\includegraphics[trim=1.5cm 10.cm 3.cm
		0cm,width=0.96\textwidth]{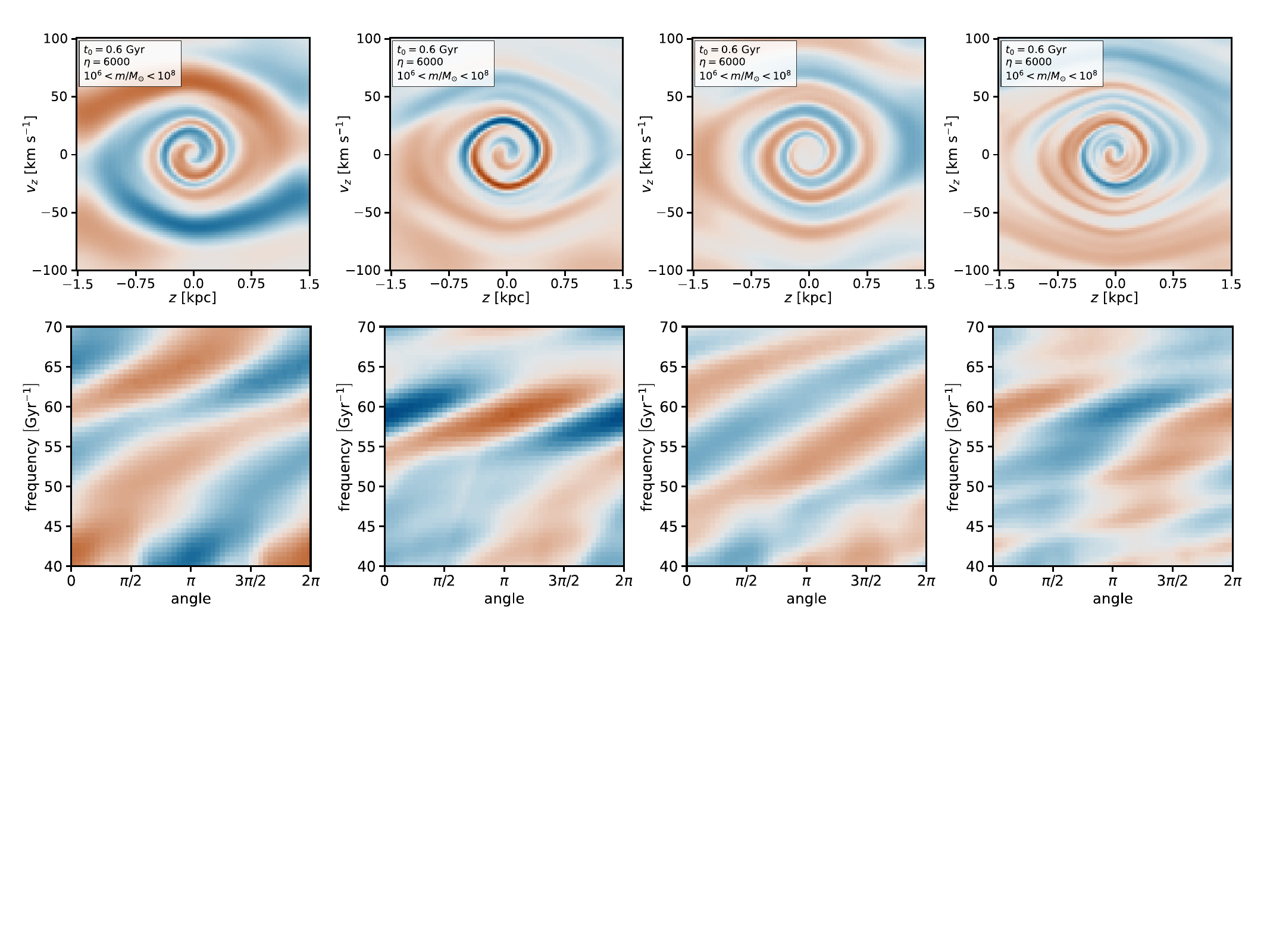}
		\caption{\label{fig:eta6000} The same as Figures \ref{fig:eta750}-\ref{fig:eta3000}, but with subhalo abundance set by $\eta=6000$.}
	\end{figure*}

\begin{figure*}
		\includegraphics[trim=1.5cm 7.5cm 3.cm
		0cm,width=0.95\textwidth]{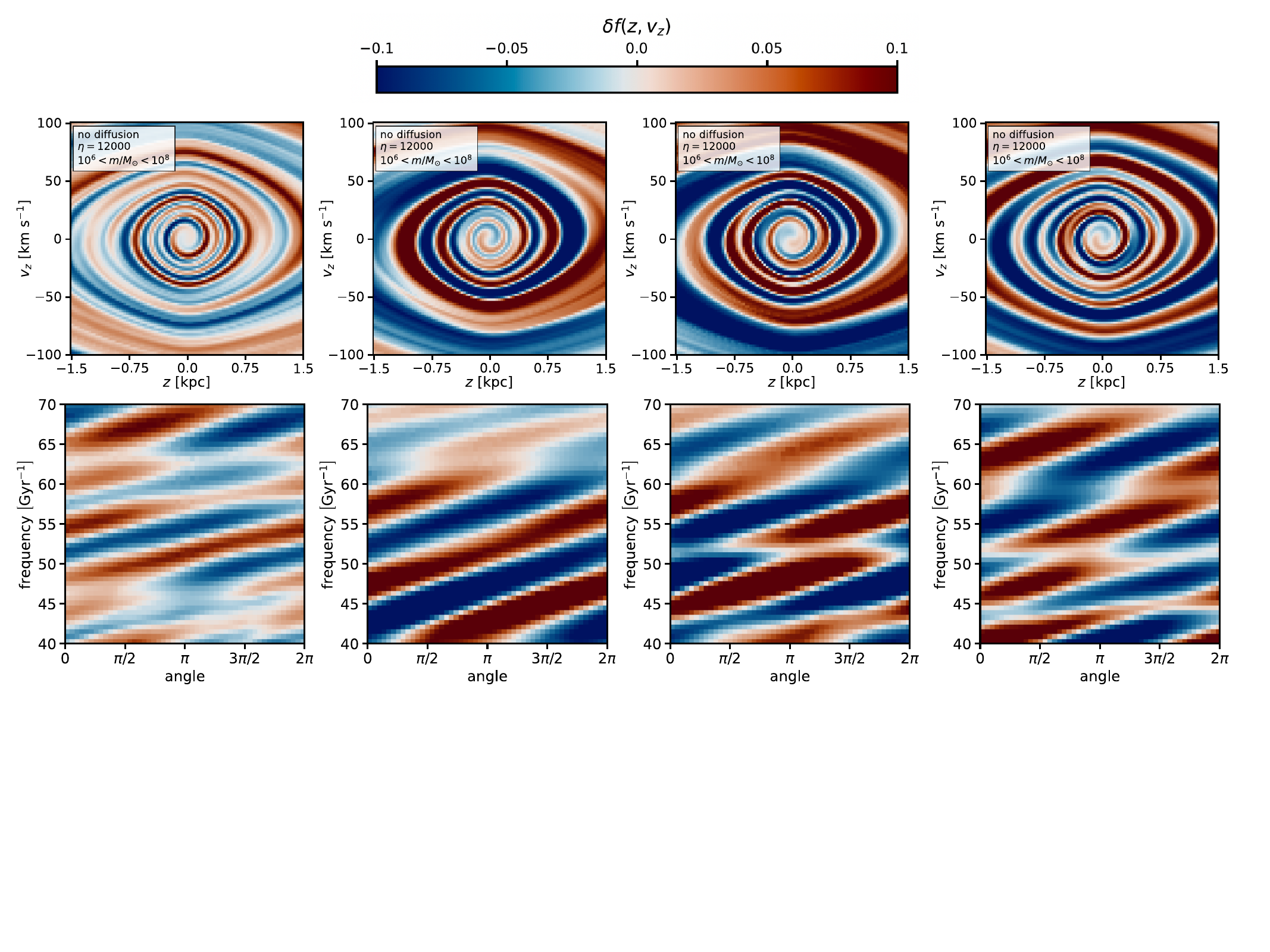}
		\includegraphics[trim=1.5cm 10.cm 3.cm
		0cm,width=0.96\textwidth]{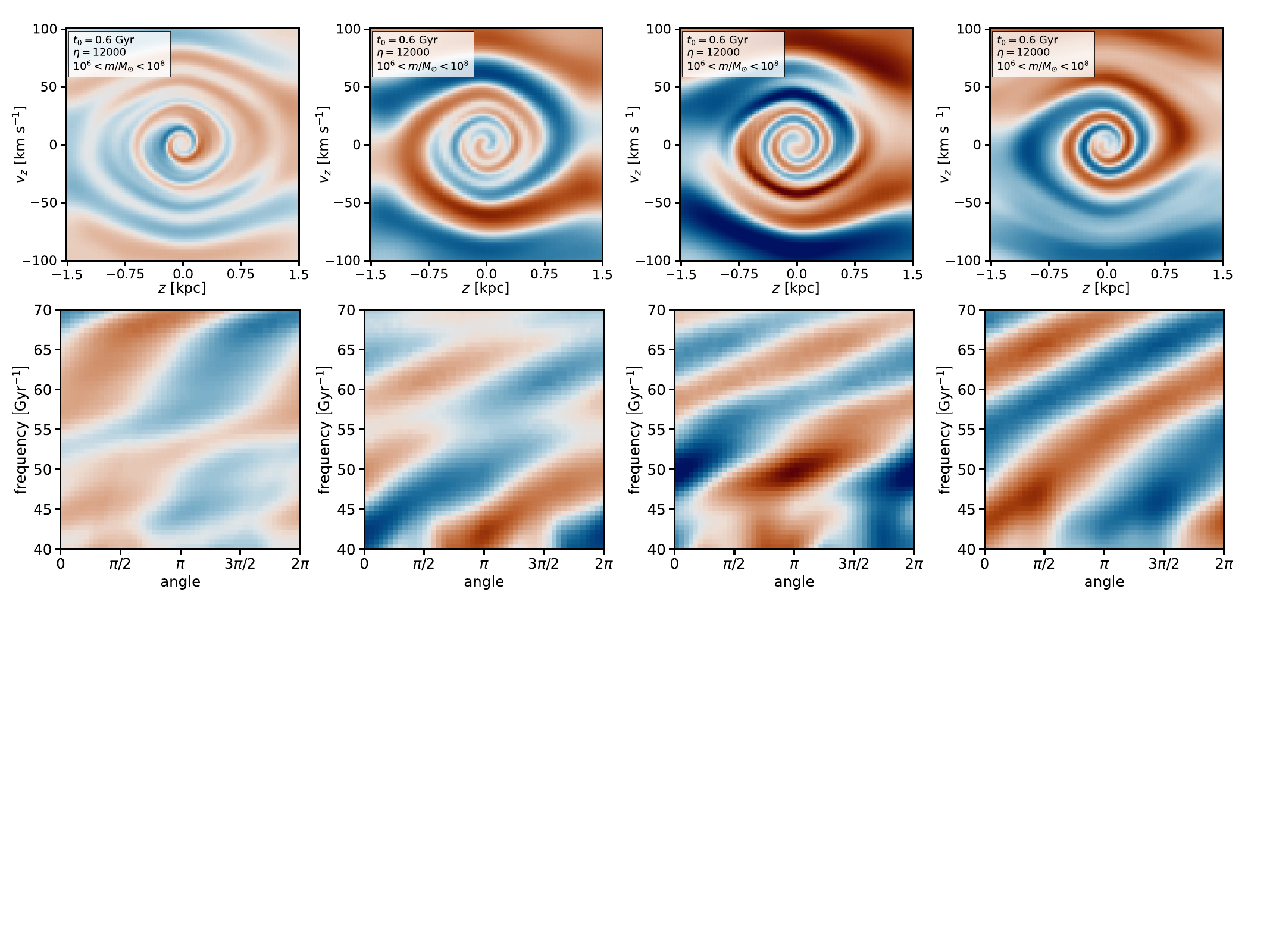}
		\caption{\label{fig:eta12000} The same as Figures \ref{fig:eta750}-\ref{fig:eta6000}, but with subhalo abundance set by $\eta=12000$.}
	\end{figure*}

	In this appendix, we provide additional visualizations of distribution functions perturbed by dark matter subhalos only with masses $10^6 < m / M_{\odot}<10^8$. The distribution functions in $\left(z,v_z\right)$ and $\left(\Omega,\theta\right)$ coordinates, with and without diffusion, are shown in Figures \ref{fig:eta750}, \ref{fig:eta1500}, \ref{fig:eta3000}, \ref{fig:eta6000}, and \ref{fig:eta12000} with subhalo abundance increasing as $\eta = 750$, 1500, 3000, 6000, and 12000, respectively. Each column depicts the same distribution function without accounting for diffusion (top two rows) and with diffusion included (bottom two rows). The first and third rows show the distribution function in $\left(z,v_z\right)$ coordinates, and the second and fourth rows show the distribution function in $\left(\Omega, \theta\right)$ coordinates. 
	
\end{document}